\documentclass[a4paper,11pt]{article}
\pdfoutput=1 

\usepackage{jcappub} 

\usepackage[T1]{fontenc} 

\usepackage[makeroom]{cancel}
\newcommand*{\QEDA}{\hfill\ensuremath{\blacksquare}}%
\usepackage[table]{xcolor}

\usepackage[normalem]{ulem} 

\usepackage{url}
\usepackage{hyperref}

\RequirePackage{doi}

\usepackage{caption}
\usepackage{subcaption}

\title{\boldmath Multimessenger Analysis Strategy for Core-Collapse Supernova Search:\\Gravitational Waves and Low-energy Neutrinos}

\author[a,*]{Odysse Halim,\note[*]{Corresponding author.}}
\author[b]{Claudio Casentini,}
\author[c,d]{Marco Drago,}
\author[e,f]{Viviana Fafone,}
\author[g]{Kate Scholberg,}
\author[h,i]{Carlo Francesco Vigorito,}
\author[j,k]{and Giulia  Pagliaroli}

\affiliation[a]{Istituto Nazionale di Fisica Nucleare (INFN) sez. di Trieste, Italy,}
\affiliation[b]{Istituto Nazionale di Astrofisica - Istituto di Astrofisica e Planetologia Spaziali (INAF - IAPS), Rome, Italy,}
\affiliation[c]{Universit\`a di Roma  La Sapienza, I-00185 Roma, Italy}
\affiliation[d]{INFN, Sezione di Roma, I-00185 Roma, Italy}
\affiliation[e]{University of Rome Tor Vergata, Rome, Italy,}
\affiliation[f]{INFN sez. di Roma Tor Vergata, Rome, Italy,}
\affiliation[g]{Department of Physics, Duke University, Durham, NC, USA,}
\affiliation[h]{University of Turin, Italy,}
\affiliation[i]{INFN sez. di Torino, Italy,}
 \affiliation[j]{Gran Sasso Science Institute (GSSI), L'Aquila, Italy,}
\affiliation[k]{INFN sez. di LNGS, Assergi, Italy}

\emailAdd{odysse.halim@ts.infn.it}
\emailAdd{claudio.casentini@gmail.com}
\emailAdd{marco.drago@gssi.it}
\emailAdd{viviana.fafone@roma2.infn.it}
\emailAdd{kate.scholberg@duke.edu}
\emailAdd{carlo.vigorito@to.infn.it}
\emailAdd{giulia.pagliaroli@gssi.it}

\abstract{Core-collapse supernovae are fascinating astrophysical objects for multimessenger studies. Gravitational waves are expected to play an important role in the supernova explosion mechanism. Unfortunately, their modeling is challenging, due to the stochastic nature of the dynamics and the vast range of possible progenitors. Therefore, the gravitational wave detection from these objects is still elusive with already advanced detectors. Low-energy neutrinos will be emitted copiously during the core-collapse explosion and can help the gravitational wave counterpart search. In this work, we develop a multimessenger strategy to look for such astrophysical objects. We exploit a global network of both low-energy neutrino and gravitational wave detectors. First, we discuss how to improve the detection potential of the neutrino sub-network by exploiting the time profile of a neutrino burst from a core-collapse supernova. We show that in the proposed approach, neutrino detectors can gain at least $10\%$ of detection efficiency at the distance where their efficiency drops. Then, we combine the information provided by gravitational wave and neutrino signals in a multimessenger analysis. { In particular, by using the clusters of low-energy neutrinos observed by LVD and KamLAND detectors in combination with the gravitational wave triggers from LIGO-Virgo detector network, we obtain an increase of the probability to detect the gravitational wave signal from a core-collapse supernova at $60$ kpc, from zero to $\sim 33\%$ for some specific gravitational wave emission model.} \\ \\ \textit{Keywords}: multimessenger, supernova, core-collapse, low-energy neutrino, gravitational wave.}

\begin{document}
\maketitle
\flushbottom

\section{Introduction\label{sec:intro}}


 
     Core-collapse Supernovae (CCSNe) are perfect astrophysical targets for multimessenger astronomy \cite{pagliaroli_PRL,leonor}. Indeed the large amount of energy produced by the stellar collapse, $\sim 10^{53}$ erg, is expected to be released as low-energy neutrinos (LENs) { with average energy around 10 MeV}, gravitational waves (GWs), and multi-wavelength electromagnetic emissions.
     
     The first neutrino detection from a CCSN in a nearby galaxy, SN1987A, observed by Kamiokande-II \cite{hirata1987}, IMB \cite{Bionta1987}, and Baksan \cite{alexeyev}, proved that CCSNe can produce a large number of MeV neutrinos which are in the sensitivity range of our detectors. Currently, there are several neutrino detectors in operation, as Super-Kamiokande \cite{superK} (Super-K), LVD \cite{lvd_det}, KamLAND \cite{kamland}, IceCube \cite{icecube}, which are sensitive to a LEN burst search to distances up to at least the edge of the Milky Way and beyond. These detectors are also involved in a joint prompt search for a CCSN neutrino burst via the SuperNova Early Warning System (SNEWS) \cite{Antonioli2004,Al_Kharusi_2021} to provide fast alerts to the electromagnetic community.
     
    The joint observation of the first binary neutron star merger \cite{Abbott2017} started the promising era of multimessenger astronomy with advanced GW detectors. The GW search is currently being carried out by the advanced detectors working as a network: two 4-km-length LIGO \cite{ligo} detectors in Hanford and Livingston, USA, and one 3-km-length Virgo \cite{virgo} detector in Cascina, Italy. These advanced detectors already performed three observing runs (O1, O2, O3) from 2015 to 2020. Moreover, the Kagra detector \cite{kagra} in Kamioka, Japan, already joined the hunt for GWs at the end of O3, with sensitivity, at a beginning stage, comparable to the other detectors.


    GWs signals are also expected from CCSN events by several different physical processes\cite{Ott_2009,Abdikamalov:2020jzn,powell2020,Szczepanczyk}.Thus, these astrophysical objects are ideal targets for multimessenger search via GWs and LENs. In this paper, we investigate the best way of combining GW and LEN data to hunt CCSNe in order to improve our efficiency and detection horizon. Here, based on our previous investigations \cite{halimtesi,halim2019}, we describe our strategy and we test its power with simulated signals injected in a time-coherent way, both in GW and LEN data.


    
    There have been several studies on multimessenger searches to combine gravitational waves and other messengers, including searches with high energy neutrinos { (TeV energy)}
    \cite{Aso_2008,multimess,Adrian2013,dipalma2014} and with gamma-ray bursts \cite{gwgrbabbott,Ashton_2018}. However, the joint analysis strategy combining LENs and GWs to hunt for CCSNe has not so far been studied thoroughly.

\begin{figure}[!ht]
\centering
\includegraphics[width=.7\linewidth]{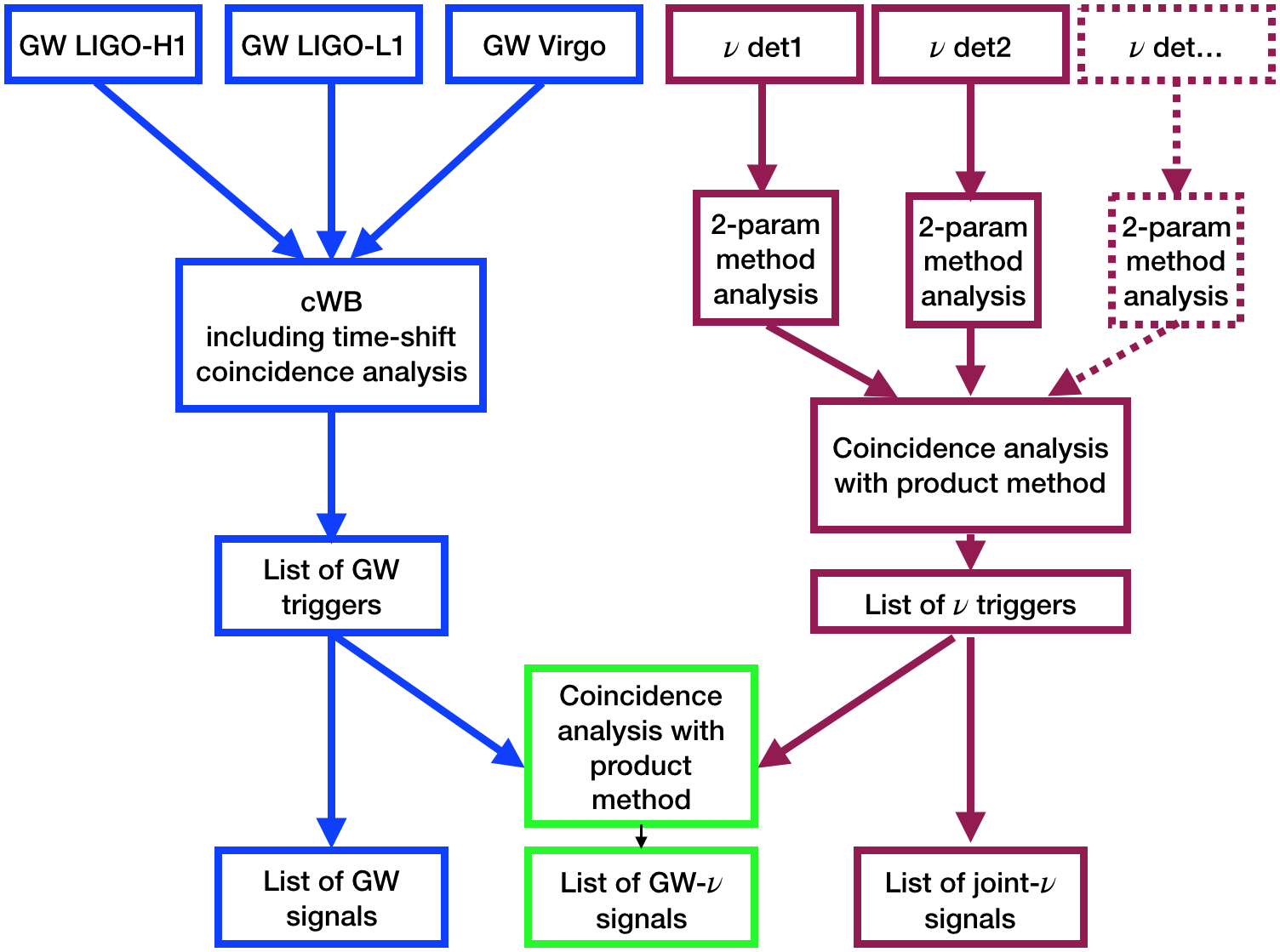}
\caption{The schematic view of the multimessenger GW-LEN strategy proposed in this paper.}
\label{fig:GWnu_scheme}
\end{figure}

    In the strategy described here, we use \texttt{coherent WaveBurst} (cWB) pipeline \cite{Klimenko_2004,Klimenko2008,dragotesi,Necula_2012} to analyze simulated GW data. This pipeline is a model-agnostic algorithm for the search of GW transients. cWB is open to a wide class of GW sources; it was the pipeline providing the first alert of the arrival of the first GW signal GW150914 \cite{gw150914} and it is used for the search of GWs from CCSNe \cite{em_gw_ccsne}. In parallel, we simulate the time series of expected LEN signal and background event rates from several neutrino detectors and then analyze the network of the simulated LEN data to hunt for astrophysical neutrinos\footnote{Note that we employ no detailed detector simulation for the neutrino detectors.}. This strategy for the neutrino network analysis will profit from a new approach, already introduced in \cite{halim2019}, to increase the burst detection sensitivity of neutrino detectors.   In this paper, we then implement a new time-coincidence analysis between two messengers, following the flow chart shown in Fig.~\ref{fig:GWnu_scheme}. Data from different messengers are analyzed separately and then combined by coincidence analysis to produce a list of possible GW-LEN signals. The described strategies could, in principle, be used for online astrophysical alert networks, such as SNEWS, or offline analysis.

The article is organised as follows. In Section \ref{sec:sources}, we will discuss the emission models from each messenger. Then, in Section \ref{sec:science}, a discussion on the data and analysis by our strategy will be presented. Finally, we will implement the  strategy on simulated data and show the result in Section \ref{sec:results}.

\section{Messengers from Core-collapse Supernovae.}
\label{sec:sources}

Several known astrophysical sources are expected to emit GWs and LENs. In this work, we consider transient sources, causing both a $O(10)$-ms GW burst  and an impulsive $O(10)$-sec emission of  $O(10)$-MeV LENs. These phenomena are expected to come from CCSNe \cite{jankarev} and ``failed'' SNe \cite{OConnor:2010moj}, which are our main focus in this article. 

{ GW and LEN signals are both sensitive to the initial conditions of CCSN simulation, as progenitor mass, rotation, etc. 
So that, a coherent combined GW-LEN analysis should be performed by considering GW and LEN signals resulting from the same numerical simulation. However, unfortunately, there are not, at the present, numerical simulations providing successful CCSN explosion and both signals. In particular, several simulations provide both signals for the first half of a second, till the explosion, which obviously is not enough to correctly estimate the neutrino emission that lasts $O(10)$-sec. 
In this paper, we made our best to relate the GW and neutrino signals coming from different available simulations with similar progenitor masses.}

\subsection{Gravitational Wave Emission}\label{sec:GWemission}

We consider the GW signals resulting from new 3D neutrino-radiation hydrodynamics CCSN simulations of Radice {\it et al.} \cite{radice} (abbreviated as ``Rad'') obtained for three different zero age main sequence (ZAMS) masses {($9\,M_\odot$, $13\,M_\odot$, and $25\, M_\odot$)} in order to take into account both low-mass progenitors with successful explosions and high-mass progenitors with failed explosions and black-hole formation. The total GW energy radiated in the different cases spans from few $10^{-11}\,M_\odot c^2$ for the lower mass of progenitor to few $10^{-9}\,M_\odot c^2$ for the $25 M_\odot$ progenitor (see Figure 4 of Radice {\it et al.}\cite{radice}). 


Moreover, we take into account also models with rapid rotation and high magnetic field. In particular, we adopt GW waveforms from two different papers, namely the Dimmelmeier model \cite{dimmelmeier2008} (abbreviated as ``Dim'') and the Scheidegger model \cite{Scheidegger:2010en} (abbreviated as ``Sch''). In this case, we use three different models from each paper with the same ZAMS mass of $15\,M_\odot$. These models produce much stronger gravitational waves. For this mechanism to work, the stellar progenitors must have strong rotation and magnetic field, which are believed to be less likely with respect to the neutrino-radiation mechanism \cite{em_gw_ccsne,jankarev,woosley}. However, we cannot rule out their existence, because we have not yet detected any CCSN GWs with any of the possible models. The amplitude evolutions of GWs are reported in c.f. Fig.~2 in \cite{dimmelmeier2008} for the Dim model and in c.f. Fig.~3 in \cite{Scheidegger:2010en} for the Sch model. The total GW energy radiated in the different cases spans from a fraction of $10^{-9}\,M_\odot c^2$ to around $10^{-7}\,M_\odot c^2$. The details of these models can be seen in Tab.~\ref{tab:gw_models}.



The adopted GW models are intended to cover as much as possible the uncertainty band on theoretical predictions, with the lower limit represented by the GW signal for the Rad model, while the upper case is the one for the Dim and the Sch models.


    \begin{table}[t]
    \caption{Waveforms from CCSN simulations used in this work. We report in the columns: emission type and reference, waveform identifier, waveform abbreviation in this manuscript, progenitor mass, angle-averaged root-sum-squared strain $h_\mathrm{rss}$, frequency at which the GW energy spectrum peaks, and emitted GW energy.}
    \label{tab:gw_models}
    \centering
    {\renewcommand{\arraystretch}{1.2}
    \begin{tabular}{|c | c | c | c | c | c | c |} 
    \hline
    Waveform & Waveform & Abbr. & Mass  & $h_\mathrm{rss}\,@10\, \mathrm{kpc}$ & $f_\mathrm{peak}$  & $E_\mathrm{GW}$\\
    Family & Identifier & & $M_\odot$ &  $\mathrm{\left[10^{-22}\,\frac{1}{\sqrt{Hz}}\right]}$ & $\mathrm{[Hz]}$ & $[10^{-9}\,M_\odot c^2]$  \\
    \hline
    \hline
        Radice \protect\cite{radice}                  & s25 & Rad25 & {25} & 0.141 & 1132 & 28 \\
      3D simulation; & s13 & Rad13 & {13}   &  0.061 & 1364  & 5.9  \\ 
        $h_+$ \& $h_\times$; (Rad)            & s9 & Rad9 & 9  & 0.031 & 460 & 0.16  \\
    \hline
    Dimmelmeier \protect\cite{dimmelmeier2008}      & dim1-s15A2O05ls & Dim1 & 15 & 1.052 & 770 & 7.685 \\
      2D simulation; & dim2-s15A2O09ls & Dim2 & 15 & 1.803 & 754 & 27.880\\
       $h_+$ only; (Dim)          & dim3-s15A3O15ls & Dim3 & 15 & 2.690  &  237 &  1.380 \\
     \hline
     
     Scheidegger \protect\cite{Scheidegger:2010en} & sch1-R1E1CA$_L$ & Sch1 & 15  & 0.129  & 1155  & 0.104  \\
    3D simulation;   & sch2-R3E1AC$_L$ & Sch2 & 15  &  5.144 & 466  & 214  \\
      $h_+$ \& $h_\times$; (Sch)         & sch3-R4E1FC$_L$ & Sch3 & 15 &   5.796  & 698  &  342 \\
     
    \hline
    \end{tabular}
    }
    \end{table}

\subsection{Low-energy Neutrino Emission}\label{sec:NUemission}

Concerning the LEN emission we consider the signals resulting from the numerical simulations of H{\"u}depohl without the collective oscillations \cite{hudepohl}.
In particular, we adopt the time-dependent neutrino luminosities and average energies obtained for a progenitor of $11.2 M_\odot$.
The simulation provides all flavors of neutrino fluxes differential in energy and time for the first 7.5 seconds of the neutrino emission; 
however, in order to cover at least the first 10 seconds of the signal, we considered also an analytical extension of these fluxes.   
The average neutrino energies from before collapse up to the simulated $0.5$ s after bounce are $\langle E_{\nu_e}\rangle=13$ MeV, $\langle E_{\bar{\nu}_e}\rangle=15$ MeV and $\langle E_{\nu_x}\rangle=14.6$ MeV, see c.f. Table 3.4 of Ref. \cite{hudepohl}.

In addition, we also adopt a parametric model for neutrino emission as described in Pagliaroli {\it et al} \cite{pagliaroli2009}. 
This model provides the best-fit emission from SN1987A data and it is characterized by a total energy radiated in neutrinos of 
$\mathcal{E}=3\times 10^{53}$ erg, and average energies of $\langle E_{\nu_e}\rangle=9$ MeV, $\langle E_{\bar{\nu}_e}\rangle=12$ MeV 
and $\langle E_{\nu_x}\rangle=16$ MeV. The temporal structure we adopt for this signal is described by:
\begin{equation}
F(t,\tau_1,\tau_2) = (1-e^{-{t / {\tau_1}}})e^{-{t / {\tau_2}}},
\label{eq:pagliaroli_model}
\end{equation}
where the parameters that govern the emission are $\tau_1$ and $\tau_2$. They represent the rise and the decay timescales of the neutrino signal. 
Their best-fit values using SN1987A data \cite{pagliaroli_ccsn} are $\sim 0.1$ s and $\sim 1$ s.

In order to simulate the clusters of supernova neutrino events we consider only the main interaction channel for water and scintillator,
i.e. the inverse beta decay (IBD) $\bar\nu_e+p \rightarrow n+e^+$.
We assume standard MSW neutrino oscillations to estimate the $\bar{\nu}_e$ flux $\Phi_{\bar{\nu}_e}$ at the detectors. 
This flux is an admixture of the unoscillated flavors fluxes at the source, 
i.e. $\Phi_{\bar{\nu}_e}=P\cdot\Phi_{\bar{\nu}_e}+(1-P)\Phi_{\bar{\nu}_x}$, 
where $x$ indicates the non-electronic flavours and $P$ is the survival probability for the $\bar{\nu}_e$. 
Depending on the neutrinos mass hierarchy, this probability can be $P\simeq 0$ for Inverted Hierarchy (IH) or $P\simeq0.7$ for Normal Hierarchy (NH).

The expected number of IBD events for the different models and detectors considered in our work is reported in Tab.~\ref{tab:nu_models} for a CCSN located at a reference distance of $10$~kpc.

    \begin{table*}[t]
    \caption{{Number of IBD events expected for a CCSN exploding at 10 kpc from us for the different neutrino models adopted and the considered detectors} (Super-K \protect\cite{superK}, LVD \protect\cite{lvd_det}, and KamLAND \protect\cite{kamland}). In parenthesis we report the assumed energy threshold ($E_\mathrm{thr}$).}
    \label{tab:nu_models}
    \centering
    {\renewcommand{\arraystretch}{1.2}
    \begin{tabular}{|c | c | c | c | c | c |} 
    \hline
    Model  & Progenitor   & Super-K & LVD  & KamLAND \\
    (identifier) & Mass & ($E_\mathrm{thr}=6.5$ MeV) & ($E_\mathrm{thr}=7$ MeV) &($E_\mathrm{thr}=1$ MeV) \\
    \hline
    \hline
    Pagliaroli \protect\cite{pagliaroli2009} & $25\, M_\odot$ & 4120 & 224 & 255\\
     (SN1987A) &&&&\\
    \hline
     H{\"u}depohl \protect\cite{hudepohl}  & $11.2\, M_\odot$ & 2620 & 142 & 154 \\
     (Hud) &&&&\\
    \hline
    \end{tabular}
    }
    \end{table*}

\section{Data and Analysis\label{sec:science}}

In this section, we will discuss the data and analysis used in our work for GWs as well as LENs. We will also present a possible strategy to do a combined multimessenger search. In the following, we assume a conservative global false alarm rate (FAR) of 1/1000 years {which is reflected in} a specific cut on FAR for the two sub-networks of LEN detectors and GW detectors; see Sec.~\ref{FAR} for a deeper discussion. 
 
\subsection{Gravitational Wave Analysis}\label{sec:gw_analysis}

The GW analysis has been done considering the cWB\footnote{cWB home page, \url{https://gwburst.gitlab.io/}; \\ public repositories, \url{https://gitlab.com/gwburst/public}\\ documentation, \url{https://gwburst.gitlab.io/documentation/latest/html/index.html}.} algorithm, a pipeline that has been widely used inside the LIGO and Virgo collaborations applied to the data of first and second generation detectors, in particular for the triggered search for CCSNe \cite{SNTargeted2016,em_gw_ccsne}. Moreover, cWB does not need any GW waveform templates; it simply combines in a coherent way the excess energy extracted from the data of the involved GW interferometers. A maximum likelihood analysis identifies the GW candidates and estimates their parameters (such as time, frequency, amplitude, etc). The candidates' detection confidence is assessed comparing the detection statistics $\rho$ with a distribution calculated from the background obtained with a time-shift procedure \cite{Klimenko:2015ypf, Drago:2020kic}.

To build the GW data set for this work, we simulate Gaussian detector noise with a spectral sensitivity based on the expected \cite{Abbott:2020qfu} Advanced LIGO and Advanced Virgo detectors \cite{ligo, virgo}. About 16 days of data have been simulated and time shifts have been performed to reach a background livetime of $\sim$ 20 years. Waveforms from emission models described in Section \ref{sec:GWemission} have been generated with discrete values of distances: 5, 15, 20, 50, 60, 700 kpc, with an incoming sky direction different for each one of them, according to the presence of possible sources. For the lower distances (5, 15, 20) we considered a Galactic model following \cite{marektesi}, whereas for the upper distances we considered fixed directions in the sky: the Large and Small Magellanic Clouds at 50 and 60 kpc respectively, and the Andromeda location at 700 kpc. These distances have been used for the multimessenger analysis with LENs, whereas intermediate distances between 60 and 700 kpc are also considered just to complete the efficiency curve\footnote{For distances between 60 and 700 kpc, we still considered the Andromeda direction, even though no known astronomical objects are present in that distance range.}. 
The injection rate is around $1/100$ per second, in order to maintain enough time difference between two consecutive waveforms. To ensure sufficient statistics, for each distance and considered model we inject around $\sim 2500$ different realizations over all the sky direction. 

\begin{figure}[!ht]
    \centering
    \includegraphics[width=.7\linewidth]{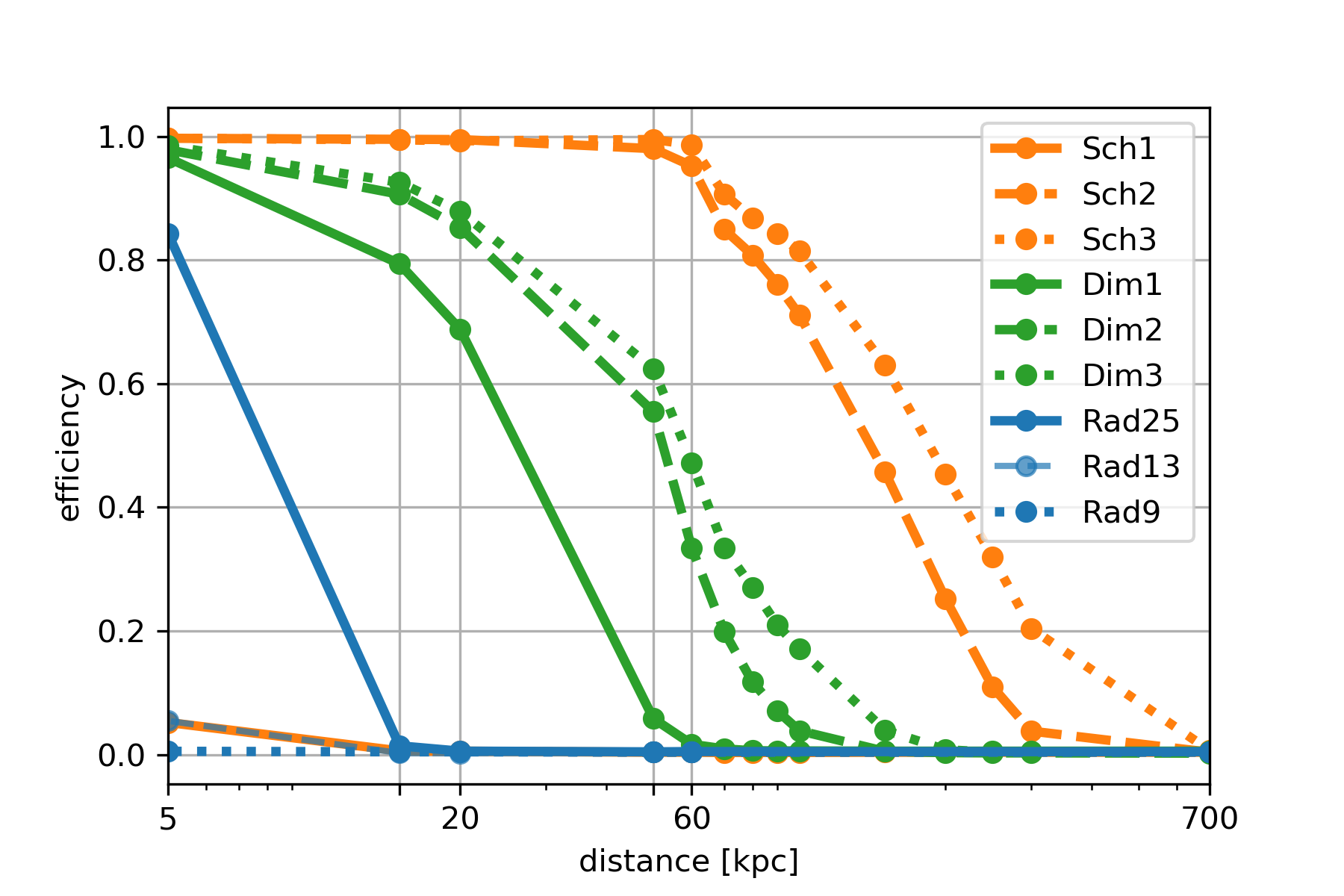} 
    \label{fig:cWBeff1}
      \caption{Efficiency curve of GW sub-network Advanced LIGO and Advanced Virgo for the different GW emission models (see Tab. \protect\ref{tab:gw_models}) and considering a FAR threshold of $864$/day.}
      \label{fig:gw_1987_60}
    \end{figure}

GW candidates are passed to the multimessenger analysis after applying a $\mathrm{FAR_{GW}}$ threshold of $864$ per day, which has been set to reach the required combined FAR of 1/1000 years. { Efficiency curves in Fig.\ref{fig:gw_1987_60} represent the ratio of the number of recovered injections with a $\mathrm{FAR_{GW}}<864$ per day to the $\sim 2500$ total ones performed for each distance.}  


\subsection{Neutrino Analysis: A New Approach to Expand the Neutrino Detection Horizon}
\label{subsec:neutrino}

In the standard LEN analysis to search for CCSNe \cite{Agafonova2014,Ikeda2007,Abe2016}, a time series data set from a detector is binned in a sliding time window of $w=20$ seconds. The group of events inside each window is defined as a \textit{cluster} and the number of events in the cluster is called multiplicity $m$. The multiplicity distribution due to background-only events is expected to follow a Poisson distribution and the significance of the $i$-th cluster is correlated with its \textit{imitation frequency} ($f^\mathrm{im}$) defined as,
\begin{equation}
f^\mathrm{im}_i (m_i)=N\times \sum_{k=m_i}^\infty P(k),
\label{eq:fim_prob}
\end{equation}
where the Poisson term, $P(k)$, represents the probability that a cluster of multiplicity $k$ is produced by the background and is defined as,           
\begin{equation}
P(k)=\frac{(f_\mathrm{bkg}w)^k e^{-f_\mathrm{bkg}w}}{k!},
\label{eq:poiss_pdf}
\end{equation}
and $N=8640$ is the total number of windows in one day, taking into account that in order to eliminate boundary problems, there is a $10$-s overlapping window between two consecutive bins. In fact, this imitation frequency is equivalent to the FAR in the GW analysis.


Based on our previous work \cite{Casentini2018} on exploiting the temporal behavior of LEN signals from CCSNe\footnote{Recent developments based on this analysis approach are also discussed in \cite{Mattiazzi:2021zcb}.}, we characterized each cluster by a novel parameter, defined as $\xi_i\equiv \frac{m_i}{\Delta t_i}$, where $\Delta t_i$ is the duration of the $i$-th cluster, i.e., the time elapsing from the first to the last event in a cluster. Obviously, in our analysis, this duration can reach a maximum value of 20 seconds, which is the bin time window size itself. Moreover we will consider only clusters with $m_i\geq2$, so that the parameter $\xi_i\geq0.1$.

Previous results \cite{Casentini2018} show that by performing an additional cut in $\xi$ it is possible to disentangle further the simulated astrophysical signals from the background. However, in this work we further investigate the possibility to use the $\xi$ parameter to define a new modified \textbf{2-parameter} ($m$ and $\xi$) imitation frequency for each cluster, called $F^\mathrm{im}$, which can be calculated as follows:
\begin{equation}
F^\mathrm{im}_i (m_i,\xi_i)=  N \times  \sum_{k=m_i}^\infty  P(k,\xi_i),
\label{eq:newfim_0}
\end{equation}
where the term $P(k,\xi_i)$ represents the joint probability that a cluster with a given multiplicity $k$ \textbf{and} a specific value of $\xi_i$ is produced by the background.

\begin{figure}[!ht]
\centering
\includegraphics[width=.7\linewidth]{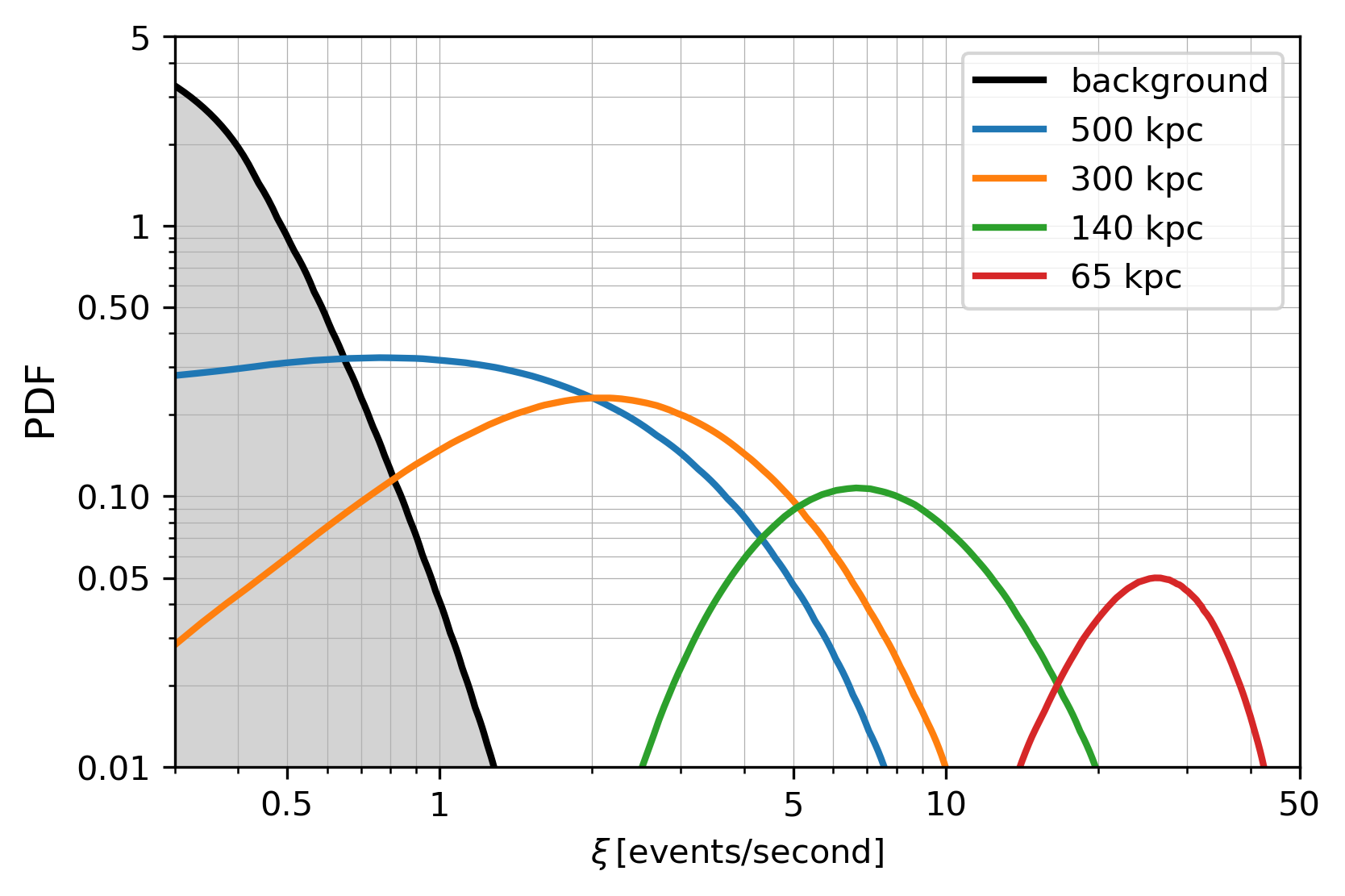}
\caption{Probability density functions for background plus signal clusters as functions of the $\xi$ parameter and for different distances in the case of the Super-K detector. The black solid line shows the PDF for pure background clusters. Data are taken from \protect\cite{Casentini2018}.}
\label{fig:pdf}
\end{figure}

It is convenient to rewrite the joint probability as $P(k,\xi)=P(\xi | k)P(k)$; where $P(\xi | k)$ is the conditional probability of a cluster to have a specific $\xi$ value given the cluster has already had a multiplicity $k$. This conditional probability can be derived for each detector by taking into account the distribution of the $\xi$ values expected for clusters only due to background \cite{Casentini2018}. 
As a leading example, we show in Figure \ref{fig:pdf} this distribution for the Super-K detector in the form of the probability density function (PDF).

Particularly in Figure \ref{fig:pdf}, the black solid line represents the normalised probability for background clusters to have a specific value of $\xi$, i.e. $\mathrm{PDF}(\xi | k)$.
Then, this $\mathrm{PDF}$ is related to the conditional probability, namely $P(\xi_i | k)= \int_{\xi=\xi_i}^\infty \mathrm{PDF}(\xi\geq\xi_i|k)\, d\xi$.  Equation \ref{eq:newfim_0} can actually be rewritten as (see App. \ref{app} and c.f. Sec.~7.1. of \cite{halimtesi} for more detail),
\begin{equation}
F^\mathrm{im}_i(m_i,\xi_i)= N \times \sum_{k=m_i}^\infty P(k) \int_{\xi=\xi_i}^\infty \mathrm{PDF}(\xi\geq\xi_i|k) d\xi.
\label{eq:newfim_1}
\end{equation}
We show in App.~\ref{app} that the new imitation frequency converges to the standard one of equation \ref{eq:fim_prob} for pure background clusters while it gives a much smaller value than the Poisson expectation for signal clusters (for larger $\xi$). 

To build the LEN data set for this work, we simulate about 10 years of background data for each neutrino detector assuming the following background frequencies $f_\mathrm{bkg}$: $0.012$ Hz for Super-K \cite{Abe2016}, $0.015$ Hz for KamLAND \cite{kam_bg}, and $0.028$ Hz for LVD \cite{Agafonova2014}. Background distributions as a function of the $\xi$ parameter are obtained from pure background data.

The CCSN simulated signals from all the emission models described in Sec.~\ref{sec:sources} are injected into the neutrino background data and \textit{coherently} into the GW background data, i.e., for each model and each source distance, the GW and LEN signals are injected so that the starting points of the two signals are coincident in time, taking also into account the temporal delay due to the different positions of the detectors. Neutrino clusters are considered signal candidates if their $f^\mathrm{im}\leq 1/$day\footnote{For the single-detector threshold, we still use the 1-parameter $f^\mathrm{im}$ described in Equation \ref{eq:fim_prob}.}, which has been set in order to reach the global FAR of $1/1000$ years. 

 \begin{figure}[!ht]
    \centering
      \includegraphics[width=.7\linewidth]{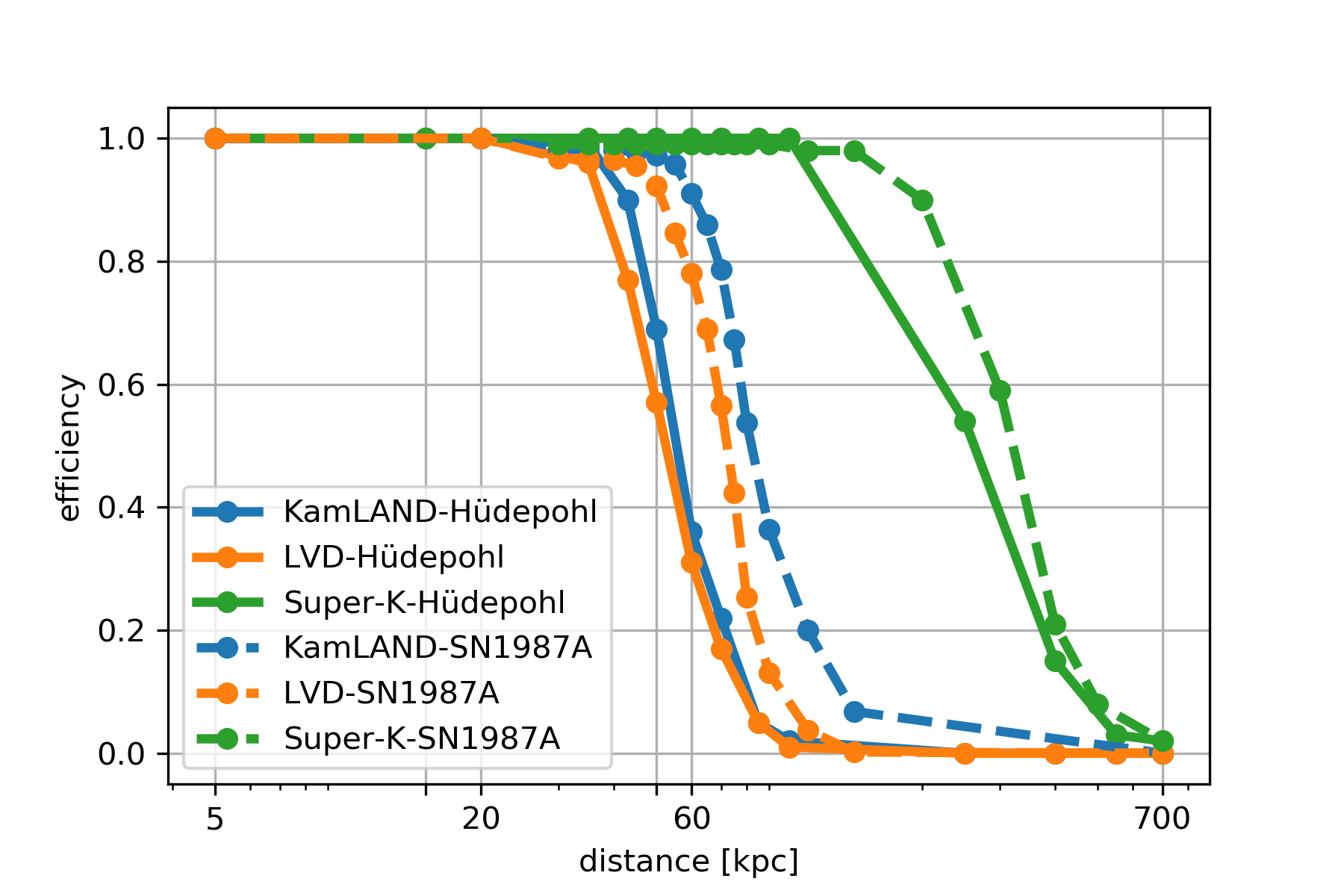}
      \caption{The efficiency curves of neutrino detectors for the Hud (continuous lines) and SN1987A emission model (dashed lines). Clusters are selected with an imitation frequency threshold of 1/day.}
      \label{fig:Efficiency}
    \end{figure}%

The LEN efficiency curves are shown in Figure \ref{fig:Efficiency} for all the detectors and emission models considered. They are defined from the simulations of signals at different distances and their subsequent injections, at a certain defined rate, into each detector's background. The cluster of expected neutrino events is extracted through a Monte Carlo and then injected into the background of each detector. After the injection, we group in clusters the output data set using the window $w$ of 20 seconds following the procedure described in previous sections. {Finally we select clusters with $f^\mathrm{im}\leq 1/\mathrm{day}$.}

In case of a network of neutrino detectors, the expected signals from the same CCSN are injected into the different-detector data sets by also taking into account the expected time of flight between the detectors. {We call this method the coherent injection}. We inject signals at a rate of 1 per day. The efficiency for each CCSN distance $D$ is defined as,
\begin{equation}
\eta(D) = \frac{N_\mathrm{r,s}(D)}{N_\mathrm{inj,s}(D)}.
\label{eq:det_eff}
\end{equation}
Here $N_\mathrm{r,s}$ is the number of recovered signals, after the selection in $f^\mathrm{im}$, while $N_\mathrm{inj,s}$ is the total number of injected signals. 



\subsection{Multimessenger Analysis}
\label{FAR}


As discussed in Sec.~\ref{sec:intro}, the ultimate goal of our analysis (green boxes in Fig.~\ref{fig:GWnu_scheme}) is to perform a multimessenger analysis, combining both neutrino triggers and GW triggers. {This is done by performing} a temporal-coincidence analysis between these two trigger lists. Joint coincidences found between LEN and GW triggers {are defined }as ``CCSN candidates''. In order to assess the statistical significance of such candidates we need to combine the FAR of GW triggers with those of LEN triggers. The $\mathrm{FAR_{GW}}$ is obtained by applying the time-shifting method described in {Sec. \ref{sec:gw_analysis}}. The $\mathrm{FAR_{\nu}}$ associated with a neutrino trigger is obtained by following the product method of $\mathrm{Nd}$-fold detector coincidence introduced in SNEWS \cite{Antonioli2004}, i.e.,
\begin{equation}
\mathrm{FAR_{\nu}}=\mathrm{Nd}\times w_{\nu}^{\mathrm{Nd}-1}\prod_{i=1}^{\mathrm{Nd}}F^\mathrm{im}_i,
    \label{eq:farLEN}
\end{equation}
where $\mathrm{Nd}$ is the number of neutrino detectors combined, $w_{\nu}$ is the time window used to look for coincidences in the neutrino sector, and $F^\mathrm{im}_i$ is the imitation frequency of the clusters obtained by exploiting the 2-parameters method described in Sec.~\ref{subsec:neutrino}.

Finally, the multimessenger {$\mathrm{FAR_{glob}}$} associated with ``CCSN candidates'' is\footnote{More discussion on the choice of coincidence analysis can be seen in \cite{halimtesi}.},
\begin{equation}
    \mathrm{FAR_{glob}}=\mathrm{Net}\times w_c^{\mathrm{Net}-1}\prod_{X=1}^\mathrm{Net}\mathrm{FAR}_X,
    \label{eq:jointfar}
\end{equation}
where $\mathrm{Net}$ is the number of sub-networks, $w_c$ is coincidence window between GW and LEN signals and $\mathrm{FAR}_X$ is the \textit{false-alarm-rate} from the sub-network $X=\{\mathrm{\nu,\,GW}\}$.

Furthermore, it is straightforward also to write the \textit{false-alarm-probability} in terms of Poisson statistics as
\begin{equation}
    \mathrm{FAP}=1-e^{-\mathrm{FAR}\times\mathrm{livetime}},
    \label{eq:jointfap}
\end{equation}
where $\mathrm{livetime}$ is the common observing time in the considered network.

By using Equation \ref{eq:jointfar} and \ref{eq:jointfap}, we can compare the performance of our 2-parameter method (Eq. \ref{eq:newfim_1}) with the standard 1-parameter method (Eq. \ref{eq:fim_prob}) in the context of multimessenger analysis. This performance can be described in terms of efficiency values, i.e., the ratio between the number of survived candidates (after all the cuts/thresholds) due to injections and the total number of injections performed.



As anticipated at the beginning of this section, we apply a threshold on $\mathrm{FAR_{glob}}$ of $1/1000$ years in order to be very conservative and a window $w_c=10$ seconds to accommodate all the scenarios. To reach the $\mathrm{FAR_{glob}}$ threshold, we impose the same requirements\footnote{Note that this requirement may change in subsequent SNEWS updates.} of SNEWS \cite{Antonioli2004} in the neutrino sub-network, which is the $\mathrm{FAR_\nu} \leq 1/100\,\mathrm{year}$ and a temporal window $w_{\nu}=10$ seconds, so that the required threshold for the GW sector is $\mathrm{FAR_{GW}}\leq 864$ per day. Let us stress that the two windows for the search for coincidences in the LEN sector and in the global network could be in principle very different. 

We define a ``detection'' in a network when $\mathrm{FAP}\geq 5\sigma$\footnote{$5\sigma\approx 5.7\times 10^{-7}$}.

 \section{Results \label{sec:results}}
 In this section, we will discuss the results {following the procedure in the previous section}. We will start by discussing the single-detector neutrino analysis results, {and then move on to} the sub-network of neutrino detectors, {and to wrap all the steps, we will provide} the global network of GW-LEN analysis.
 
 

 \subsection{Improving the LEN detection capability}\label{sec:single_det}
 
\begin{figure}[!ht]
    \centering
      \includegraphics[width=.7\linewidth]{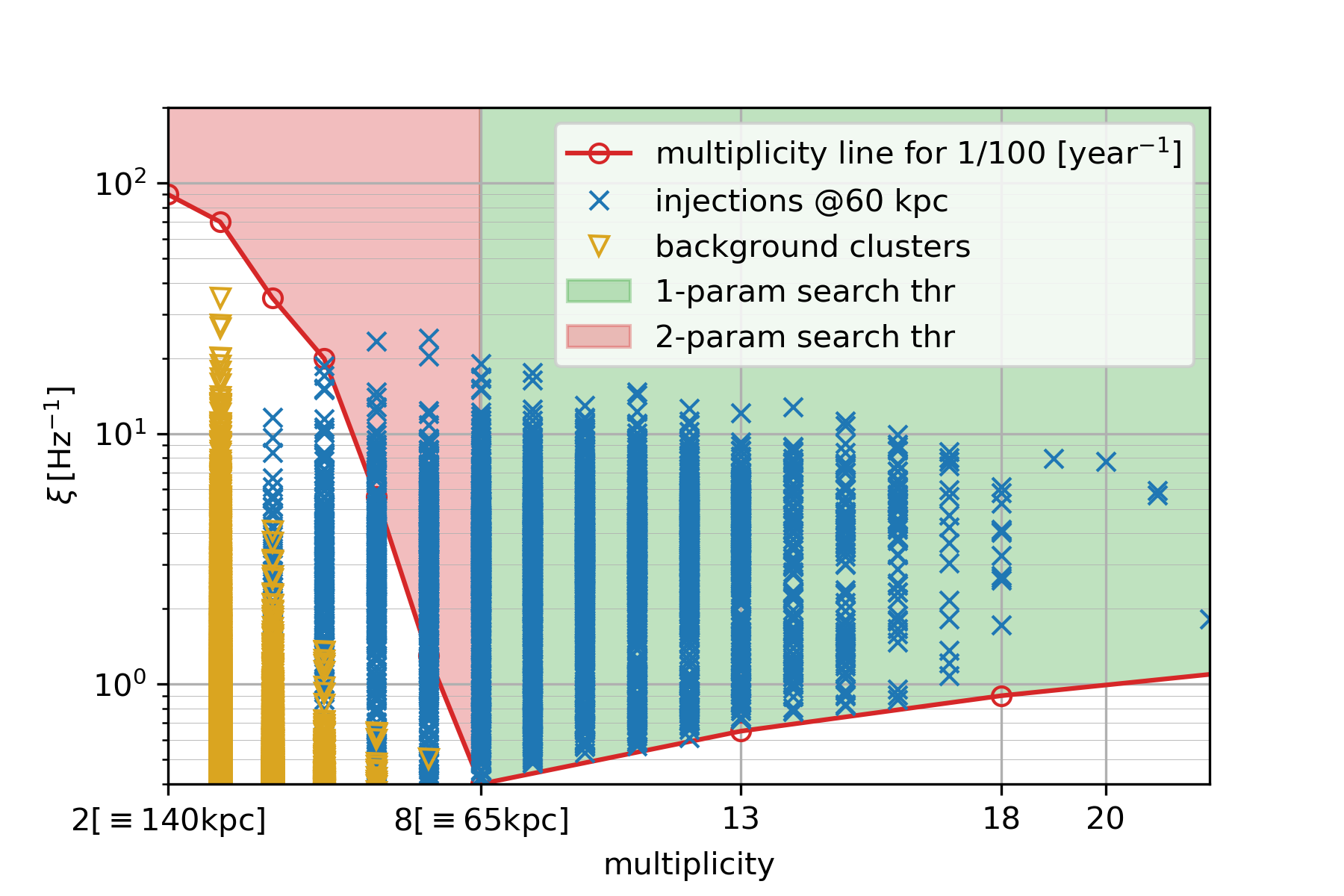}
      \caption{The $\xi$-multiplicity map for KamLAND (as a leading example) with the simulated backgrounds (yellow triangles) and injections (blue crosses) with the SN1987A emission model at 60 kpc.}
      \label{fig:kamland_60kpc}
    \end{figure}

 We apply our method to analyze simulated single detector data for KamLAND, LVD, and Super-K taking into account both neutrino emission models described in Section \ref{sec:NUemission}. 
 In order to quantify the improvement related to the 2-parameter method versus the standard one, we discuss in the following, as a leading example, the case of the KamLAND detector. 
 Let us consider a CCSN occurring at 60 kpc with the neutrino signals following SN1987A model (see the first row of Tab \ref{tab:nu_models}). After simulating {10 years} of KamLAND background data we inject randomly these simulated signals with the rate of 1 per day (3650 in total). All clusters reconstructed by the analysis are plotted in Fig. \ref{fig:kamland_60kpc} in a $\xi$ {vs} multiplicity plane. 
 { Each blue cross in this plot represents one cluster of events generated by one injected CCSN signal. The cluster multiplicity of the injections can be different despite the CCSN distance is fixed to 60 kpc. The reason is that the Monte Carlo simulation automatically allows the statistical Poisson fluctuation of the IBD events, moreover the number of background events inside the 20 second window is also fluctuating. For each cluster we estimate the associated imitation frequency (or $\mathrm{FAR}_\nu$\footnote{As previously stated in Sec. \ref{subsec:neutrino}, the imitation frequency can be considered as the $\mathrm{FAR}$. Thus, in this case, $\mathrm{FAR}_\nu$ is basically imitation frequency for KamLAND data, either $f^\mathrm{im}$ or $F^\mathrm{im}$ depending on the context in the text.}). This imitation frequency in standard 1-parameter analysis is only one-to-one related to the cluster multiplicity through Poisson statistics. So that, in order to fulfill the SNEWS requirement of $\mathrm{FAR_\nu}\leq1/100$ year, a cluster's multiplicity needs to be at least equal to 8. }
 
    
{ In other words the cluster should lie on the green area of Fig.\ref{fig:kamland_60kpc} and all the injected clusters with a multiplicity $<8$ are lost by standard 1-parameter analysis. This lower limit on the multiplicity could be translated into a maximum KamLAND horizon of $\simeq 65$ kpc for the emission model based on SN1987A, indeed the average multiplicity expected for a CCSN at this distance is $\langle m_i\rangle=8$.

With our 2-parameter method, the imitation frequency of each cluster is also a function of the $\xi$ value following Eq. \ref{eq:newfim_0} and we can determine the pair of $\xi$ and multiplicity values in order to have the needed $\mathrm{FAR_\nu}$. In particular, the red line in Fig. \ref{fig:kamland_60kpc}  belongs to $\mathrm{FAR_{\nu}}=1/100$ years for the 2-parameter method, i.e., the threshold corresponding to the current SNEWS requirement. All the clusters above this red line fulfill the $\mathrm{FAR_{\nu}}<1/100$ years requirement and this happens also with multiplicity lower than 8 given a specific $\xi$ value. The red area in Fig.~\ref{fig:kamland_60kpc} represents the improvement area, i.e., clusters which pass the FAR threshold for 2-parameter method but \textbf{not} for the 1-parameter. 
In addition, we show that all simulated background clusters (yellow triangles) are well below the red threshold line.

The result of Fig. \ref{fig:kamland_60kpc} could be interpreted as an increase of the efficiency as quantified in Tab.~\ref{tab:kam_single}. The efficiency for KamLAND to identify a signal at 60 kpc is improved from $73\%$ in 1-parameter method to $83\%$ by using 2-parameter method. Moreover, this result also implies that we are expanding the detection horizon of the detector.}




\begin{table}[!ht]
    \caption{Efficiency ($\eta$) comparison between 1-parameter and 2-parameter methods for the single detector KamLAND at 60-kpc for $\mathrm{FAR_{\nu}}<1/100\,\mathrm{[year^{-1}]}$ with the SN1987A model.} 
    \label{tab:kam_single}
    \centering
    {\renewcommand{\arraystretch}{1.2}
    \begin{tabular}{|c | c | c | c |} 
    \hline
    Noise & Noise   & $\eta_\mathrm{1param}$ & $\eta_\mathrm{2param}$  \\
     & $\left[<1/100\,\mathrm{yr}\right]$ &  $\left[<1/100\,\mathrm{yr}\right]$ & $\left[<1/100\,\mathrm{yr}\right]$  \\
     \hline
    \hline
    75198 & 0/75198 & \cellcolor{magenta!30}2665/3654=\textbf{72.9\%}  & \cellcolor{yellow!70}3026/3654=\textbf{82.8\%}   \\
    \hline
    \end{tabular}
    }
    \end{table}

 The results obtained for KamLAND in this specific case are representative for all the scenarios we investigated. Similar improvements can be obtained for different emission models and various neutrino detectors. Details and figures are in App.~\ref{appC} for the Super-K detector analysis.
 



\subsection{The sub-network of LEN detectors}
In this section, we extend the analysis to the sub-network of LEN detectors.
As stated previously, we consider several neutrino detectors in this work and we can construct sub-networks of pair configurations. {Our aim is to show the impact of the 2-parameter method in this specific sector of the analysis}. Thus, we will discuss the combined analysis of KamLAND and LVD detectors, given that their efficiency curves are very similar; see Fig. \ref{fig:Efficiency}.

\begin{figure}[!ht]
    \centering
      \includegraphics[width=.7\linewidth]{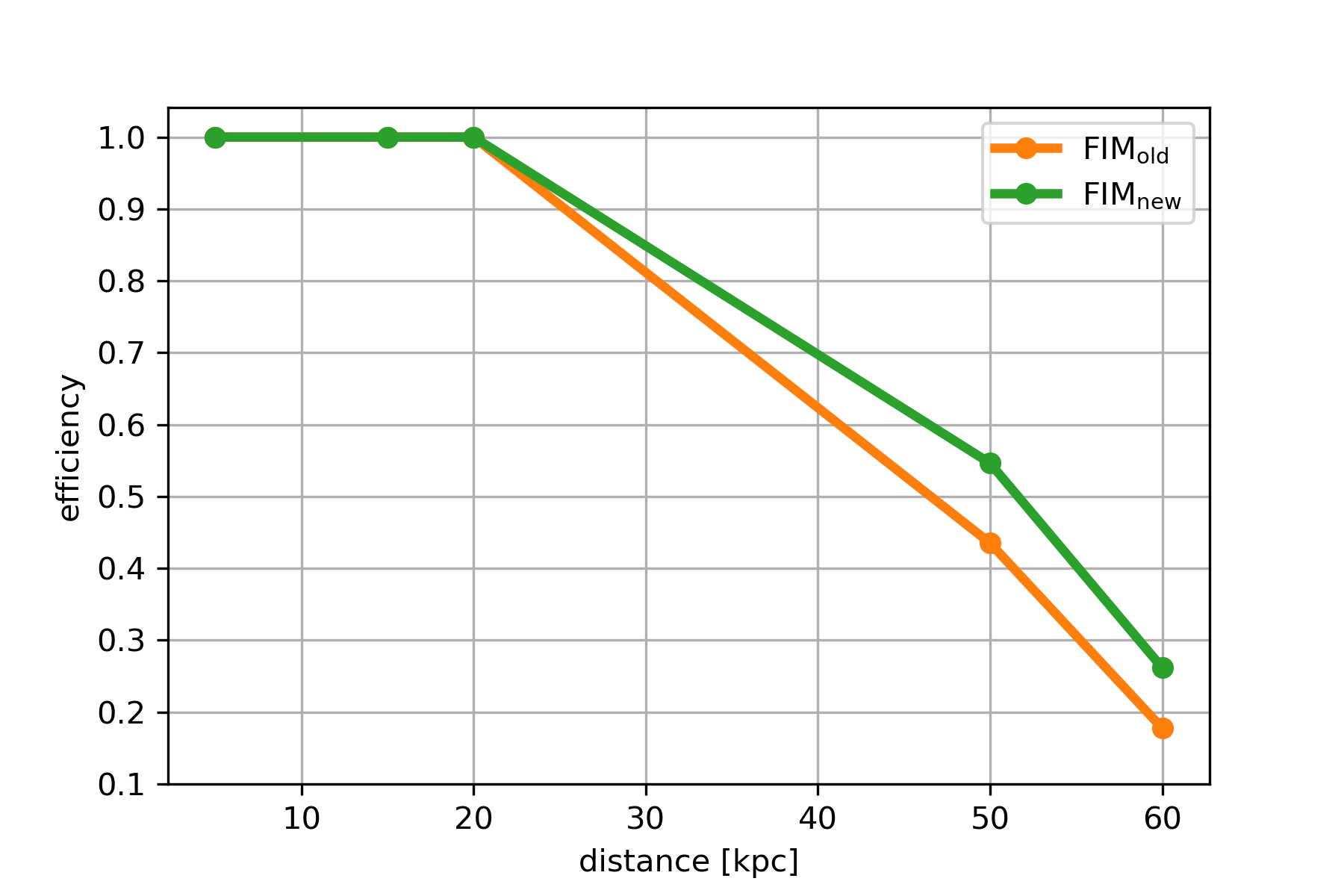}
      \caption{The efficiency of KamLAND-LVD analysis from the Hud model when a threshold on $\mathrm{FAP_\nu}\geq5\sigma$ is applied to the network. The orange line is the one obtained with the old 1-parameter method while the green line shows the improvement of the new 2-parameter method presented in this paper.}
      \label{fig:efficiecny_double}
    \end{figure}

Let us consider the neutrino signal from the Hud model and CCSNe {happening} at 5, 15, 20, 50, 60, and 700 kpc from us. Injections are performed {{in a coherent way (Section \ref{subsec:neutrino})}} in both data sets and coincidences in time within {$w_\nu=10$} sec are considered as potential CCSNe signals. In this sub-network of neutrino detectors, we put a threshold of $5\sigma$ in $\mathrm{FAP}_\nu$ (Eq. \ref{eq:jointfap}). 
We compare in Fig. \ref{fig:efficiecny_double} the efficiencies at this threshold for the 1-parameter method (orange line) and the 2-parameter one (green line), and for the point of interest, we show more details of this comparison for 50 and 60 kpc in Tab. ~\ref{tab:kam_lvd_eff}.

 The efficiency to identify these signals at a distance of 50 kpc with a $\mathrm{FAR_\nu}\le 1/100$ years is $12\%$ and $26\%$ for LVD and KamLAND, respectively. However, if the detectors work together looking for time coincidences within $w_\nu$, the number of recovered signals above the same statistical threshold grows to $\sim 43\%$ when adopting the standard SNEWS requirment for the FAR estimation (the 1-parameter method). Finally, when also the $\xi$ value is taken into account (2-parameter method) this efficiency grows to $\sim 55\%$. Following the same logic, for a CCSN at 60 kpc, the fraction of signals with $\mathrm{FAR_\nu}\le 1/100$ years is only $3\%$ and $7\%$ for LVD and KamLAND, respectively. When they work as a network this efficiency increases to $18\%$ with the standard FAR estimation and to $26\%$ with the new method proposed in this work.





\begin{table}
    \caption{Efficiency ($\eta$) comparison between 1-parameter and 2-parameter method for analysis of KamLAND-LVD with the Hud neutrino model and for $\mathrm{FAP_\nu}>5\sigma$.}
    \label{tab:kam_lvd_eff}
    \centering
    {\renewcommand{\arraystretch}{1.2}
    \begin{tabular}{|c | c | c |} 
    \hline
    Distance $\mathrm{[kpc]}$ &  $\eta_\mathrm{1param}$ & $\eta_\mathrm{2param}$ \\
     & $\left[>5 \sigma \right]$ &  $\left[> 5 \sigma\right]$ \\
    \hline
     \hline
     50 & \cellcolor{magenta!30}\textbf{47/108=43.5\%} & \cellcolor{yellow!70}\textbf{59/108=54.6\%}\\
    \hline
     60 &  \cellcolor{magenta!30}\textbf{19/107=17.8\%} & \cellcolor{yellow!70}\textbf{28/107=26.2\%}\\
    \hline
    \end{tabular}
    }
    \end{table}


It is important to clarify that the efficiency curves in Fig.~\ref{fig:efficiecny_double} cannot be directly compared with the ones reported in Fig.~\ref{fig:Efficiency} for the single detector cases, because the requirement applied in terms of $\mathrm{FAR_{\nu}}$ is different. In order to help the reader compare the results, we briefly report the results also for the SN1987A emission model and a CCSN at $60$ kpc detected with the LVD-KamLAND network. The increase in efficiency is from {$85\%$ to $93\%$} and these numbers can be directly compared with the ones in Tab. \ref{tab:kam_single} related to the single detector case.




\subsection{The global network of GW-LEN detector}

After discussing the joint-neutrino analysis, here we discuss the global analysis in order to combine neutrino and gravitational waves in a single network. { In the global GW-LEN network we look for temporal coincidences within $w_c=10$ seconds among GW triggers and LEN clusters. We assess their statistical significance following the approach discussed in Sec. 2. 
In this paper, we would like to emphasize the power of combining GWs and LENs in a situation where both sub-networks of detectors can gain on combining data. In other words, we highlight the case in which the detection efficiency of both LEN and GW detectors is not $100\%$ to see the improvement in both directions. To do this, we need to combine detectors with similar detection efficiency at the same CCSN distance. 

As reported in previous sections, the horizon of the GW network is completely dependent on the assumed GW emission model, see Fig. \ref{fig:gw_1987_60}. In particular, for the model called Dim2 in Tab. \ref{tab:gw_models}, the GW detection horizon is compatible with that of the LVD and KamLAND neutrino detectors and coincides with the Large Magellanic Cloud. 
For this reason, we highlight the results for the global network LIGO-Virgo, LVD, and Kamland.}
Results for different GW models are reported in App.~\ref{appB}. 

In this section, we adopt as the requirement in terms of significance $5\sigma$ to claim real GW-$\nu$ detection. In particular, we show, for all the networks and emission models considered, the $\mathrm{FAR_{glob}}$ defined in Eq. \ref{eq:jointfar} of ``CCSN candidates'' estimated by following the standard 1-parameter method procedure (we call this $\mathrm{FAR_{old}}$) versus the same quantity obtained with the new 2-parameter method (called $\mathrm{FAR_{new}}$).
\begin{figure}[!ht]
    \centering
      \includegraphics[width=.7\linewidth]{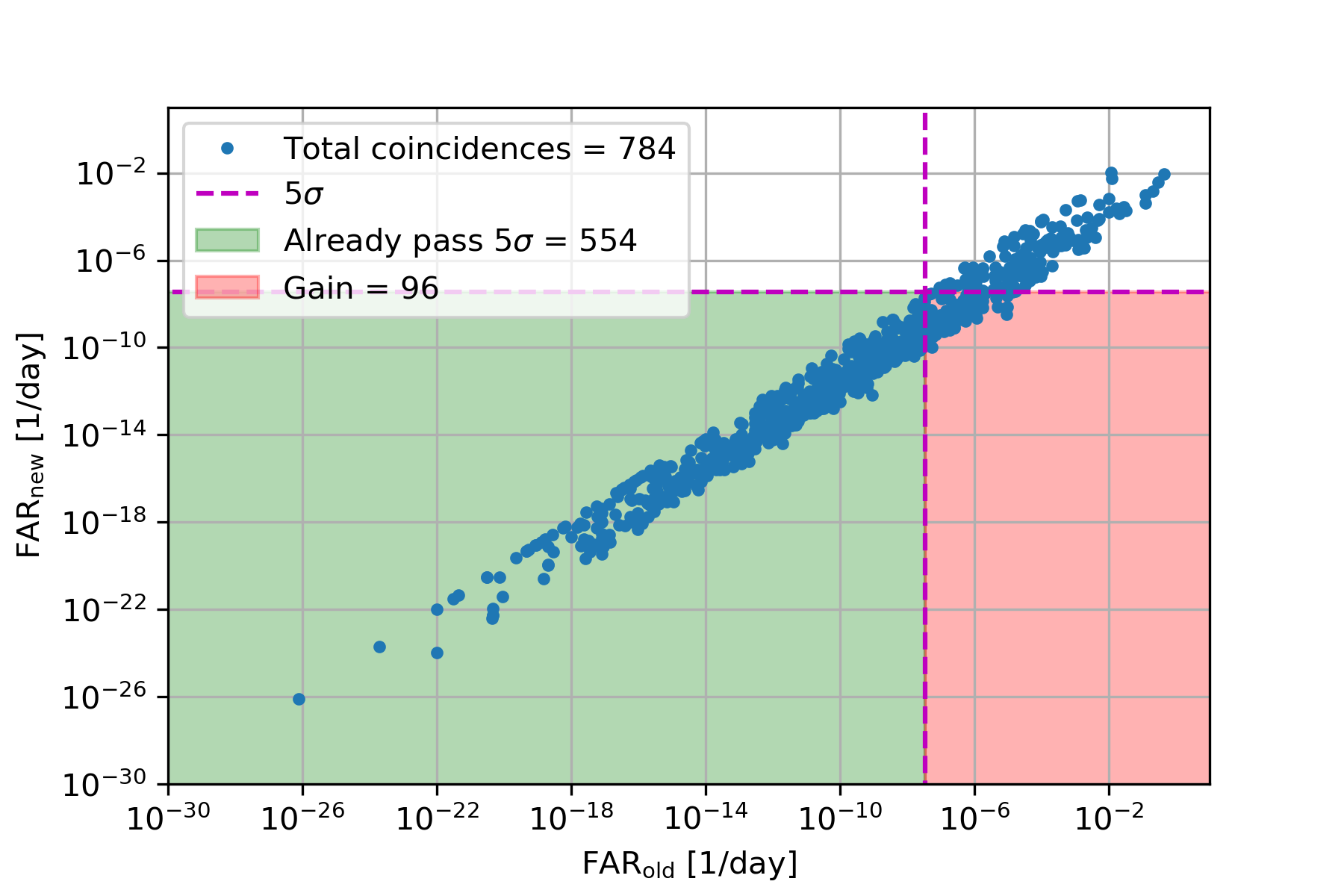}
      \caption{{The $\mathrm{FAR_{glob}}$ of GW-$\nu$ candidates obtained with the 2-parameter method ($\mathrm{FAR_{new}}$) vs the 1-parameter ($\mathrm{FAR_{old}}$) considering KamLAND (SN1987A-model) and HLV (Dim2-model) and a CCSN at 60 kpc.}}
      \label{fig:gw_1987_60_dou}
    \end{figure}%
    
Now we discuss the case of KamLAND detector working together with the HLV (Hanford, Livingston, Virgo) GW network. In Fig.~\ref{fig:gw_1987_60_dou} we compare the $\mathrm{FAR_{old}}$ with the $\mathrm{FAR_{new}}$ for the case of a CCSN occurring at $60$ kpc and with a neutrino emission compatible with SN1987A and the GW emission Dim2. The magenta dashed line corresponding $5\sigma$ significance is also plotted. The green area is the area where the 1-parameter method produces $5\sigma$ significant clusters. Meanwhile, the red area is the improvement region produced by 2-parameter method. The blue data points belonging to the red zone are CCSN signals detected only with the new procedure. As reported in the first line of Tab.~\ref{tab:dimkamlvd} the 2-parameter method gives us an additional {$\sim12\%$} of signals that otherwise are lost by 1-parameter method. 


     \begin{figure}[!ht]
    \centering
    \includegraphics[width=.7\linewidth]{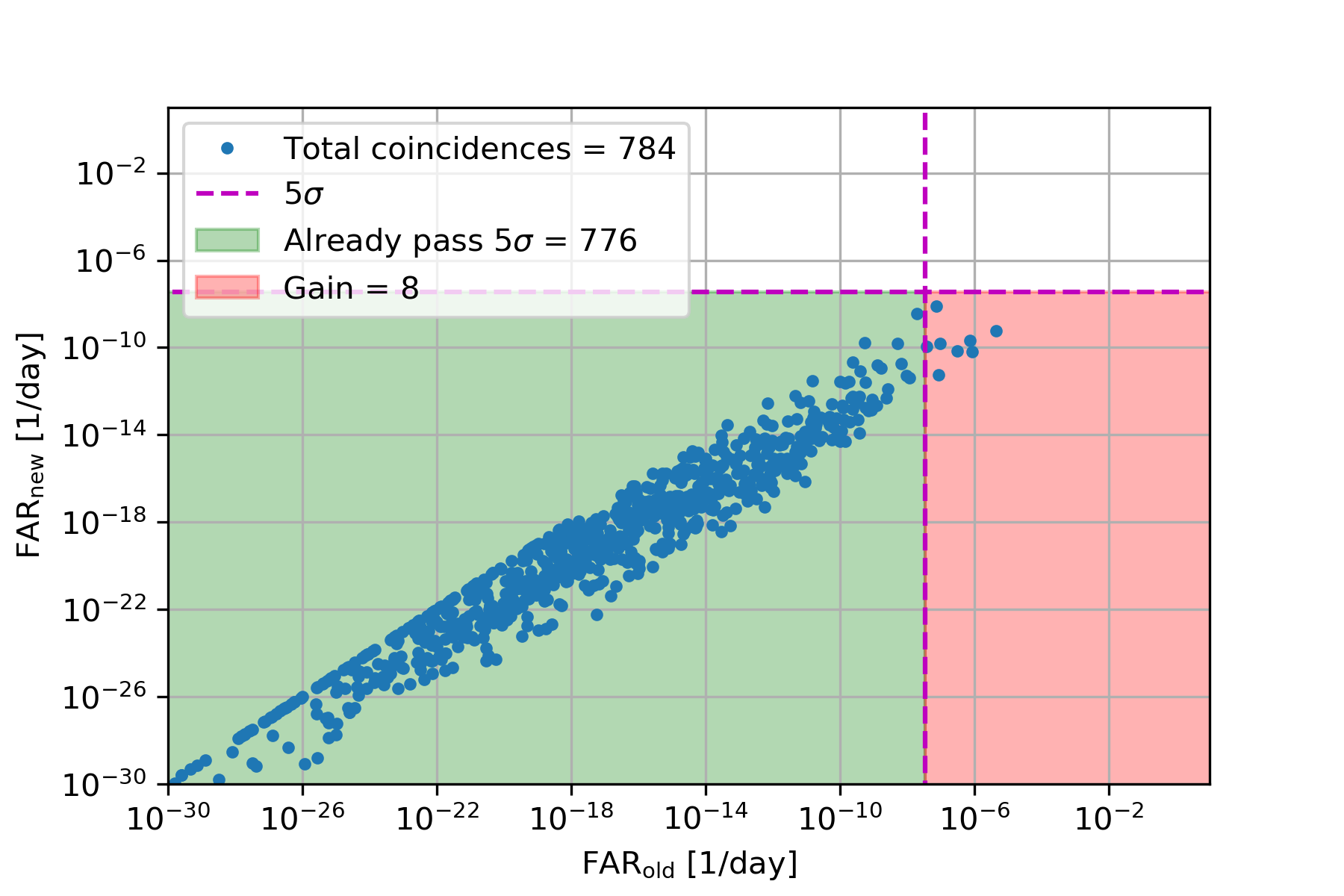}
      \caption{{The $\mathrm{FAR_{glob}}$ of GW-$\nu$ candidates obtained with the 2-parameter method ($\mathrm{FAR_{new}}$) vs the 1-parameter ($\mathrm{FAR_{old}}$) considering KamLAND-LVD (SN1987A-model) and HLV (Dim2-model) for a CCSN at 60 kpc.}}
      \label{fig:gw_1987_60_tri}
    \end{figure}%

\begin{table}
    \caption{Efficiency ($\eta$) comparison of 1-parameter and our 2-parameter method for Figure \protect\ref{fig:gw_1987_60_dou} and \protect\ref{fig:gw_1987_60_tri}. The first column indicates the specific network of detectors considered and the adopted emission models. The second column shows results after we impose the threshold on the FAR of GW data ($\mathrm{<864/day}$). The third and last columns report the fraction of signals with a significance greater than $5\sigma$ (efficiency) with 1-parameter and 2-parameter methods.}
    \label{tab:dimkamlvd}
    \centering
    {\renewcommand{\arraystretch}{1.2}
    \begin{tabular}{|c | c | c | c |} 
    \hline
    Network $\&$ Type  & Recovered   & $\eta_\mathrm{1param}$ & $\eta_\mathrm{2param}$  \\
    of Injections & $\mathrm{FAR_{GW}}<864/\mathrm{d}$ &  $\left[>5\sigma\right]$ & $\left[>5\sigma\right]$  \\
    \hline
    \hline
    HLV-KAM  & 784/2346= &  554/784= &  650/784=  \\
    (Dim2-SN1987A) & 33.4\% & \cellcolor{magenta!30}\textbf{70.7\%} &  \cellcolor{yellow!70}\textbf{82.9\%} \\
    \hline
    HLV-KAM-LVD  & 784/2346= & 776/784= &  784/784=   \\
    (Dim2-SN1987A) & 33.4\% & \cellcolor{magenta!30}\textbf{99.0\%}  &  \cellcolor{yellow!70}\textbf{100\%}  \\
    \hline
    \end{tabular}
    }
    \end{table}


In other words, 2346 GW injections are performed and analyzed with the cWB GW-pipeline, 784 of them show a $\mathrm{FAR_{GW}<864/day}$. These GW triggers are considered to look for temporal coincidences with the list of neutrino clusters characterized by {$\mathrm{FAR_{glob}}<1/1000$} year, preliminary. We eventually chose to use a $5\sigma$ threshold in order to compare the results with Ch.~8 of c.f. Ref.~\cite{halimtesi}. Among the candidates GW-LEN coincidences, {554} have a statistical significance of $5\sigma$ ($\sim 71\%$ of the GW triggers) when the standard 1-parameter method is used. By applying the new 2-parameter method, we gain about 110 more signals detected increasing the fraction of GW triggers to $\sim 83\%$. 


Following the same approach, we extend this result by considering also the LVD detector inside the neutrino sub-network. Results in this case are reported in Fig.~\ref{fig:gw_1987_60_tri} and in the second row of Tab.~\ref{tab:dimkamlvd}. In this case, the improvement due to the 2-parameter method versus the standard one seems less evident.  The reason is that the efficiency saturates to its maximum value of $33.4\%$, i.e., all the GW triggers are in coincidence with a neutrino candidates whose significance is $>5\sigma$.

\begin{figure}[!ht]
    \centering
      \includegraphics[width=.7\linewidth]{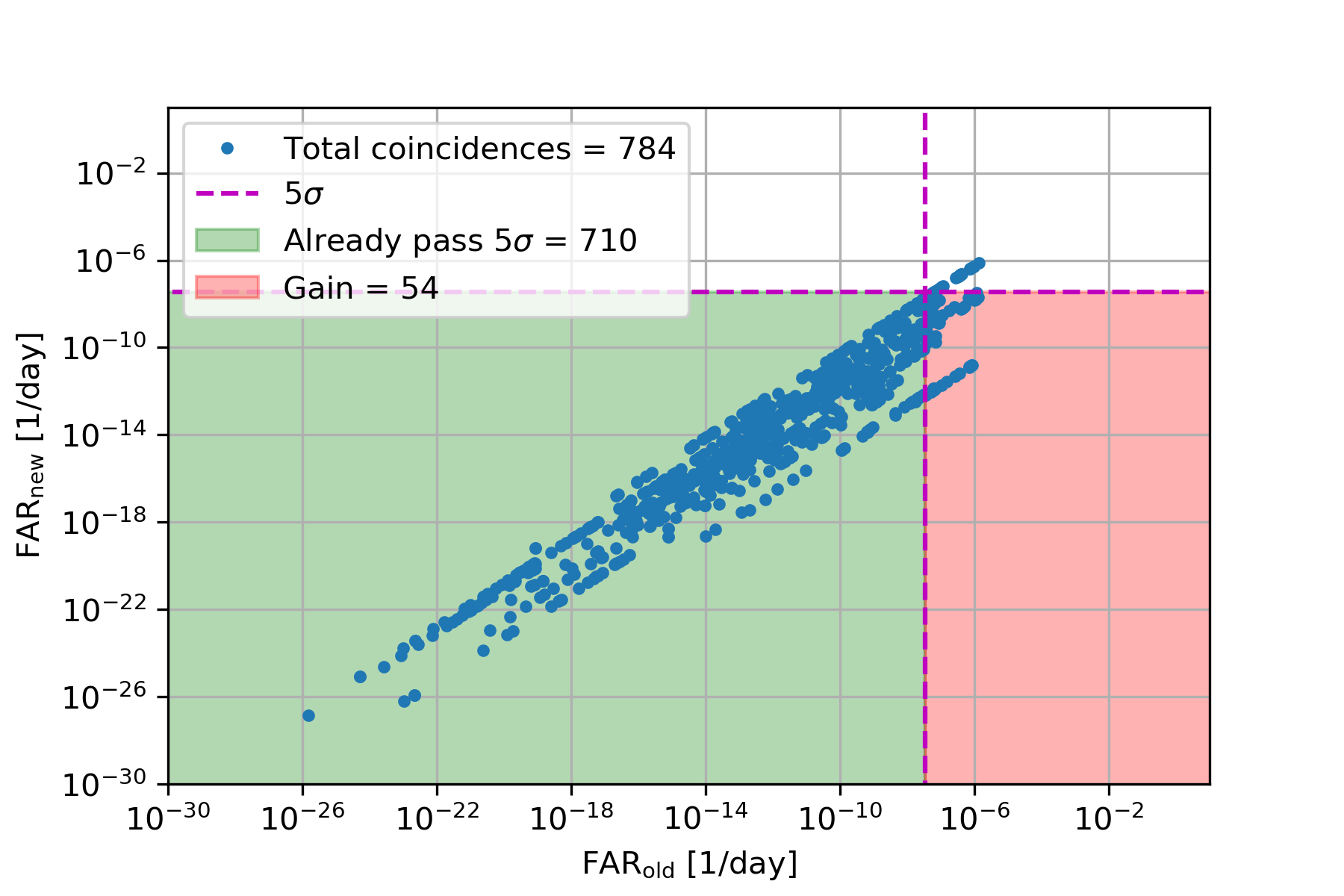}
      \caption{{The $\mathrm{FAR_{glob}}$ comparison of our 2-parameter method ($\mathrm{FAR_{new}}$)  vs the 1-parameter ($\mathrm{FAR_{old}}$) with coincidence analysis between joint neutrino KamLAND-LVD (Hud-model) and GW injections (Dim2-model) at 60 kpc.}}
      \label{fig:dim2_no_60}
    \end{figure}

For the sake of completeness, we discuss also a similar case by adopting the Hud model for a CCSN occurring at 60 kpc. The comparison of the $\mathrm{FAR_{new}}$ with the $\mathrm{FAR_{old}}$ can be seen in Fig.~\ref{fig:dim2_no_60}. Moreover, the efficiency comparison on this analysis can be seen in detail in Tab.~\ref{tab:dimkamlvd_hud}, where we have $\sim7\%$ improvement with our 2-parameter method. The gain we obtain is on average $O(10^3)$ between the 1-parameter and the 2-parameter method for both emission models.

    \begin{table}
    \caption{Efficiency ($\eta$) comparison of 1-parameter and our 2-parameter method for Figure \protect\ref{fig:dim2_no_60}. The columns are analogous to Table \protect\ref{tab:dimkamlvd}.}
    \label{tab:dimkamlvd_hud}
    \centering
    {\renewcommand{\arraystretch}{1.2}
    \begin{tabular}{|c | c | c | c |} 
    \hline
    Network $\&$ Type & Recovered   & $\eta_\mathrm{1param}$ & $\eta_\mathrm{2param}$  \\
    of Injections & $\mathrm{FAR_{GW}}<864/\mathrm{d}$ &  $\left[>5\sigma\right]$ & $\left[>5\sigma\right]$  \\
    \hline
    \hline
    HLV-KAM-LVD & 784/2346=  & 710/784=  & 764/784=    \\
    (Dim2-Hud) & 33.4\%  & \cellcolor{magenta!30}\textbf{90.6\%} & \cellcolor{yellow!70}\textbf{97.5\%}  \\
    \hline
    \end{tabular}
    }
    \end{table}

Finally, let us summarize the results presented and their interpretation in terms of detection efficiency. In Fig.~\ref{fig:gw_1987_60} we can see that the GW network HLV, by applying a threshold $\mathrm{FAR_{GW}}\le 864$/day, recovers about {$\sim33\%$} of the injected signals for a CCSN distance of 60 kpc for the Dim2 GW emission model. Moreover, these recovered GW triggers are far from statistically significant; indeed by requiring a $5\sigma$ threshold, the HLV detection efficiency drops to zero. Correspondingly, for the neutrino network LVD+KamLAND, the recovered signal at {$\mathrm{FAP}_\nu>5\sigma$} is about $26\%$ \& $85\%$ for Hud and 1987A model. We note that once the threshold is set for the two sub-networks, the lower one among the two efficiencies represents the upper limit for the global network. By working as a global network and by using our method, the global detection efficiency of the GW-$\nu$ grows to $\sim 33\%$. In case of a weaker neutrino emission such as the one of Hud model, the detection efficiency of the GW-$\nu$ network reaches the value of {$33.4\%\cdot97.5\%=32.6\%$}.  


  %

\section{Conclusion \label{sec:conclusion}}

We have discussed a new multi-messenger strategy with GWs and {LENs} in order to catch signals from core-collapse supernovae. The strategy involves several LEN detectors as well as GW detectors. We considered different emission models both for GW and for LEN resulting from recent numerical simulations. We performed a coherent set of injections by taking into account also the different backgrounds characterizing the detectors, and we analyzed the detection efficiency of the global network to the different signals and for several detectors configurations.
We showed that in general a multi-messenger approach can give better sensitivity to otherwise statistically unimportant signals.


Additionally we improved the neutrino analysis sector by introducing a new parameter $\xi$ that changes the estimation of {FAR} as well as the significance value for event clusters in neutrino detectors. Thanks to this 2-parameter method, we have shown that we can get a promising improvement in terms of FARs of recovered injections without misidentification of noise. This 2-parameter method increases the detection horizon of current-generation neutrino detectors.  This approach can be easily applied also in an online system such as SNEWS2.0, which will gain in term of safe alerts for the electromagnetic community that would otherwise be lost.




{Moreover, the multimessenger campaign between GWs and LENs will profit this new method in terms of the global  efficiency gain. Due to the fact that the efficiency of LEN analysis is somewhat higher than that of GW analysis, we may in the future detect a coincidence from several LEN detectors. In this case, we can do a targeted search of GWs and we can still profit from this new method.}

\section{Acknowledgement \label{sec:acknowledgement}}

We thank Jade Powell for pointing us the supernova repository. 
The authors gratefully acknowledge the support of the NSF for the provision of computational resources. The work of GP is partially supported by the research grant number 2017W4HA7S ``NAT-NET:
Neutrino and Astroparticle Theory Network'' under the program PRIN 2017 funded by the Italian Ministero dell'Istruzione, dell'Universita' e della Ricerca (MIUR).

\appendix
\section{Derivation of equation \ref{eq:newfim_1}}
\label{app}
$\int_{\xi=\xi_\mathrm{min}}^\infty \mathrm{PDF}(\xi\geq\xi_\mathrm{min}|k)$ and the integration can be stated as,
\begin{equation}
\begin{split}
1&\equiv\int_{\xi=\xi_\mathrm{min}}^\infty \mathrm{PDF}(\xi\geq\xi_\mathrm{min}|k) d\xi\\
&=\int_{\xi_\mathrm{min}}^\infty N_kf(\xi)d\xi\\
&=\int_{k/w}^\infty N_kf(\xi)d\xi\\
&=N_k \int_{k/w}^\infty f(\xi)d\xi.
\end{split}
\label{eq:pdf_func}
\end{equation}
Then, from equation \ref{eq:pdf_func}, the normalization factor $N_k$ can be written as,
\begin{equation}
N_k=\frac{1}{\int_{k/w}^\infty f(\xi)d\xi}.
\end{equation}

The conditional probability (integral) in equation \ref{eq:newfim_1}, can be written,
\begin{equation}
\begin{split}
P(\xi | k) & =  \int_{\xi\geq k/w}^\infty N_k f(\xi) d\xi\\
& =1-\int_{k/w}^{\xi\geq k/w}  N_k f(\xi) d\xi\\
& = 1- N_k\int_{k/w}^{\xi\geq k/w}   f(\xi) d\xi\\
&= 1-\frac{\int_{k/w}^{\xi\geq k/w}   f(\xi) d\xi}{\int_{k/w}^\infty f(\xi)d\xi},
\label{eq:cond_pro}
\end{split}
\end{equation}
where the integration in the numerator always has this relation $\xi\geq k/w$, with $w$ as the maximum duration, which is the window or bin width itself.

All in all, after considering equation \ref{eq:newfim_1} to \ref{eq:cond_pro}, the new imitation frequency becomes,
\begin{equation}
\begin{split}
F^\mathrm{im}_i & (w,m_i,\xi_i) = 8640\times \sum_{k=m_i}^\infty P(k)\left[1-N_k\int_{k/w}^{\xi_i}f(\xi)d\xi\right]\\
= & f^\mathrm{im}_i(w,m_i) - 8640\times\sum\limits_{k=m_i}^\infty P(k)N_{k}\int_{k/w}^{\xi_i} f(\xi)d\xi\\
 = &  f^\mathrm{im}_i(w,m_i)-8640\times \sum\limits_{k=m_i}^{m_i+n;\,n\leq (w\cdot\xi_i-m_i)} P(k)N_{k}\\
 &\times\int_{k/w}^{\xi_i} f(\xi)d\xi.
 \label{eq:newfim_final}
\end{split}
\end{equation}

Let us test this formula. Intuitively, we can say that when we have a pure background cluster with multiplicity $m_\mathrm{bkg}$ and $\xi_\mathrm{bkg}=m_\mathrm{bkg}/w$, the new imitation frequency should be very similar as the old one, namely,
\begin{equation}
F^\mathrm{im}_\mathrm{bkg}(w,m_\mathrm{bkg},\xi_\mathrm{bkg})\simeq f^\mathrm{im}_\mathrm{bkg}(w,m_\mathrm{bkg}),
\label{eq:test_min}
\end{equation}
and when we have a very strong signal, with $m_\mathrm{strong}$ and $\xi_\mathrm{strong}$,
\begin{equation}
F^\mathrm{im}_\mathrm{strong}(w,m_\mathrm{strong},\xi_\mathrm{strong})\ll f^\mathrm{im}_\mathrm{strong}(w,m_\mathrm{strong}).
\label{eq:test_max}
\end{equation}

We can prove those relations above. First, suppose that we found a cluster whose $\xi_i=\xi_{min}=m_i/w$, meaning that $n=0$ for equation \ref{eq:newfim_final}, and thus, our new imitation frequency becomes the old one,
\begin{equation}
\begin{split}
F^\mathrm{im}_i(w,& m_i,\xi_\mathrm{min}=m_i/w)=  \left[f^\mathrm{im}_i-8640 \times\sum\limits_{k=m_i}^{m_i+0} P(k)N_{k}\right.\\
&\left.\times\int_{k/w}^{m_i/w} f(\xi)d\xi \right]\\
=&  \left[f^\mathrm{im}_i-8640 \times P(m_i)N_{m_i}\cancelto{\boxed{=0}}{\int_{m_i/w}^{m_i/w} f(\xi)d\xi} \right]\\
= & f^\mathrm{im}_i
\end{split}
\end{equation}
meanwhile, even when $\xi$ large, this condition that $N_{k}\int_{k/w}^{m_i/w} f(\xi)d\xi\leq 1$ is true, and,
\begin{equation}
\begin{split}
F^\mathrm{im}_i(w,m_i,&\xi_\mathrm{large})=  f^\mathrm{im}_i(w,m_i)-8640 \times\sum\limits_{k=m_i}^{m_i+n} P(k)N_{k}\\
&\times\int_{k/w}^{\xi_\mathrm{large}} f(\xi)d\xi \\
< &f^\mathrm{im}_i(w,m_i)-8640\times\sum\limits_{k=m_i}^{m_i+n} P(k)\\
<&8640\times\sum_{k=m_i}^\infty P(k)-8640\times\sum\limits_{k=m_i}^{m_i+n} P(k)\\
< & 8640\times \left[\left(1-\sum_{k=0}^{m_i-1} P(k)\right)-\sum\limits_{k=m_i}^{m_i+n} P(k)\right]\\
<&  8640\times \left(1-\sum_{k=0}^{m_i+n} P(k)\right)\\
<&8640\times \left(\sum_{k=m_i+n+1}^\infty P(k) \right)=f^\mathrm{im}_i(w,m_i+n+1),
\end{split}
\end{equation}
and when $n\gg$, this relation holds $f^\mathrm{im}(w,m_i+n+1)\ll f^\mathrm{im}(w,m_i)$, and therefore,

\begin{equation}
\begin{split}
F^\mathrm{im}_i(w,m_i,\xi_\mathrm{large})&<f^\mathrm{im}(w,m_i+n+1)\\
&\ll f^\mathrm{im}(w,m_i)
\end{split}
\end{equation}
Thus, we have proven the relations \ref{eq:test_min} and \ref{eq:test_max}.$\QEDA$

\section{More GW models}
\label{appB}
\begin{figure}[!ht] 
  \begin{minipage}[b]{0.5\linewidth}
    \centering
    \includegraphics[width=\linewidth]{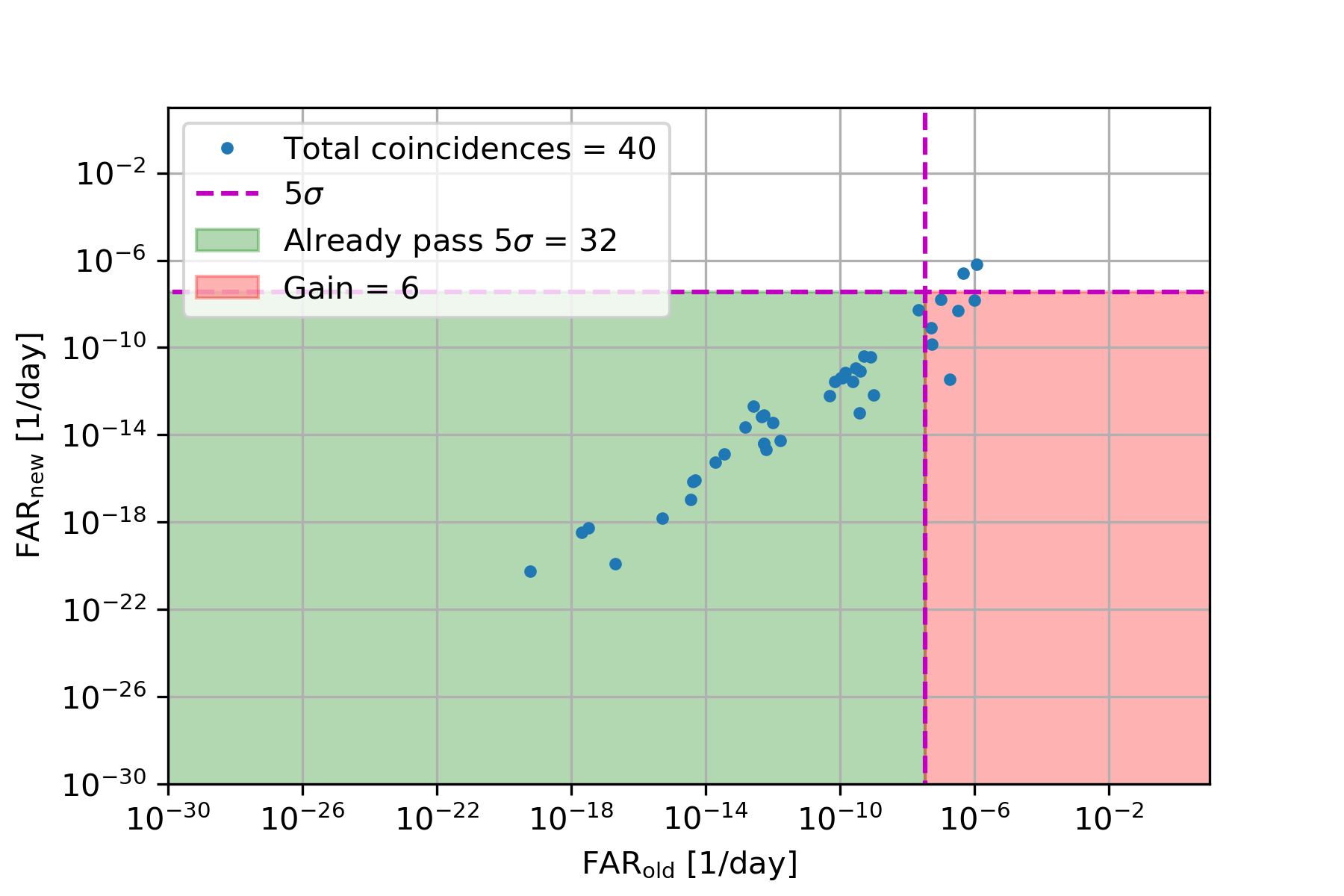} 
    \subcaption{Dim1} 
  \end{minipage}
  \begin{minipage}[b]{0.5\linewidth}
    \centering
    \includegraphics[width=\linewidth]{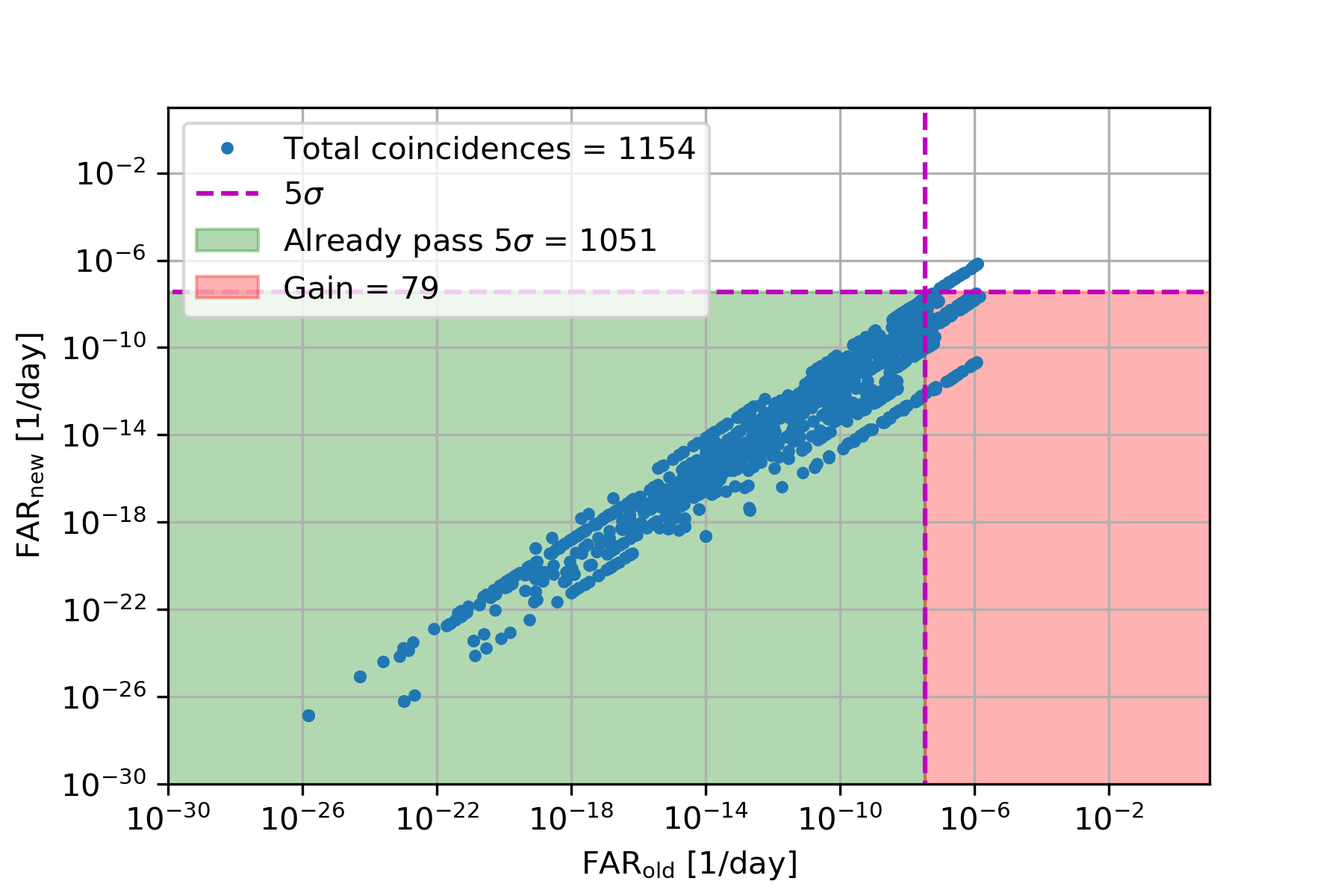} 
    \subcaption{Dim3} 
  \end{minipage} 
  \begin{minipage}[b]{0.5\linewidth}
    \centering
    \includegraphics[width=\linewidth]{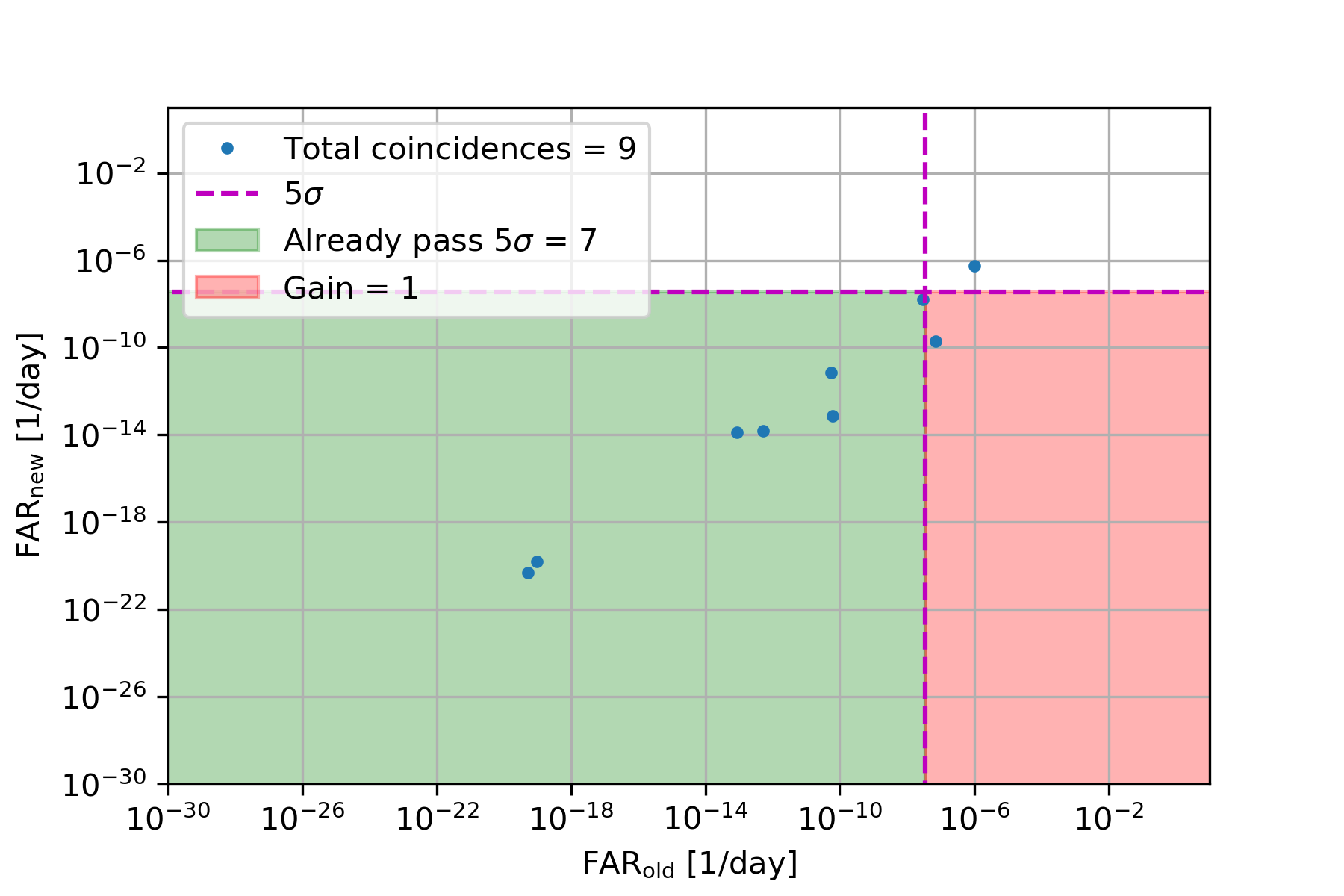} 
    \subcaption{Sch1} 
  \end{minipage}
  \begin{minipage}[b]{0.5\linewidth}
    \centering
    \includegraphics[width=\linewidth]{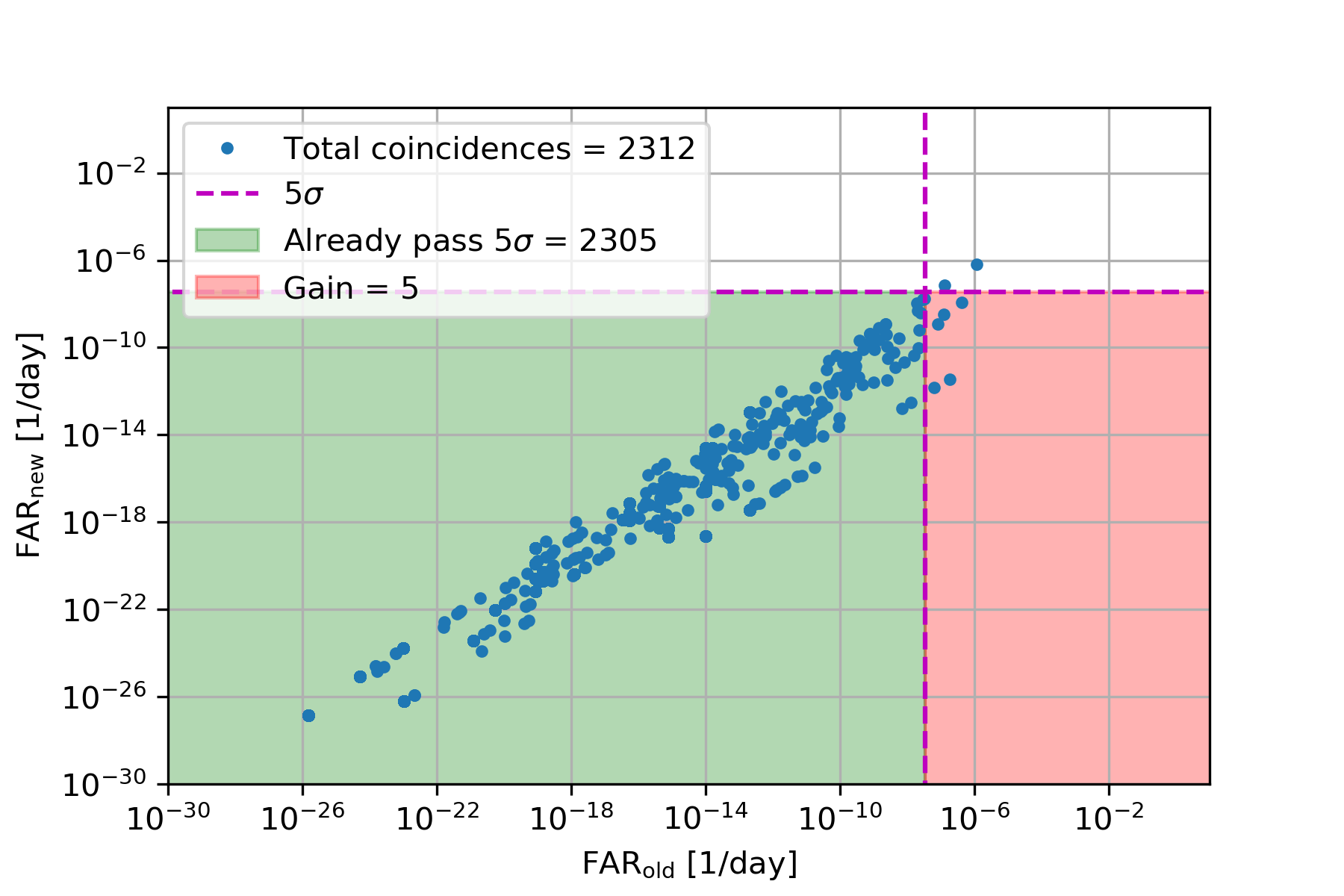} 
    \subcaption{Sch2} 
  \end{minipage}
  \centerline{
  \begin{minipage}[b]{0.5\linewidth}
    \centering
    \includegraphics[width=\linewidth]{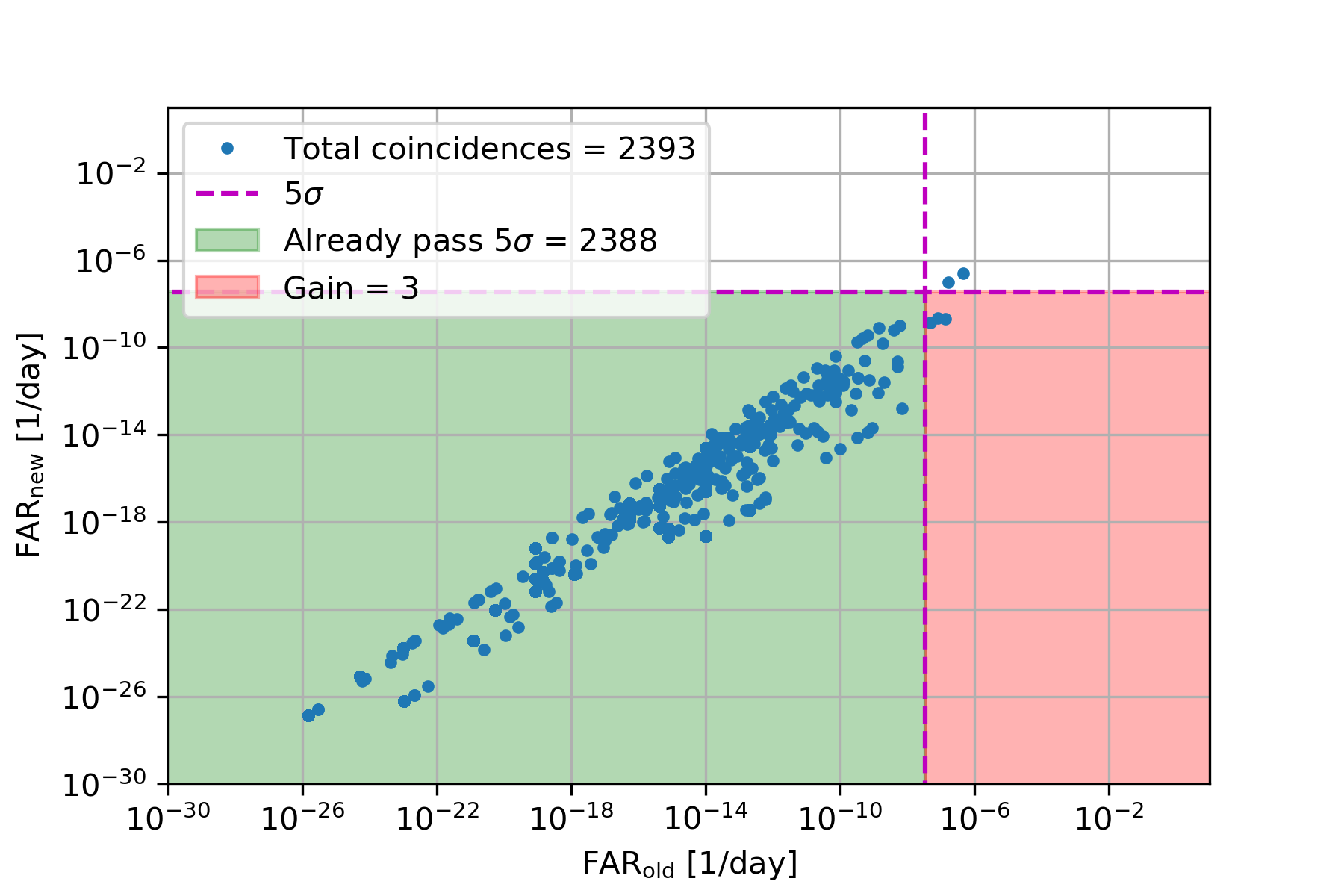} 
    \subcaption{Sch3} 
  \end{minipage}
  }
  \caption{\textbf{The $\mathrm{FAR_{glob}}$ comparison of our 2-parameter and 1-parameter method for the triple-coincidence analysis (KamLAND-LVD-GW) with injections at 60 kpc. We compare various GW models (see Table \ref{tab:gw_models}).}} 
  \label{fig:jointfig} 
\end{figure}

Here, we perform our study for other GW models such as Dim1, Dim3, Sch1, Sch2, and Sch3. All the FAR comparisons from these models can be seen in Fig.~\ref{fig:jointfig}. Dim1 and Sch1 are quite weak models, so very few injections are recovered. Meanwhile, Dim2 (in Figure \ref{fig:dim2_no_60}), Dim3, Sch2, and Sch3 are quite strong models, so we have more recovered injections after the triple coincidence analysis, and indeed more passed the $5\sigma$ significance.

    \begin{table*}
    \caption{Efficiency ($\eta$) comparison of 1-parameter and our 2-parameter method for Figure \protect\ref{fig:jointfig}. The columns are analogous to Table \protect\ref{tab:dimkamlvd}.}
    \label{tab:varkamlvd}
    \centering
    {\renewcommand{\arraystretch}{1.2}
    \begin{tabular}{|c | c | c | c |} 
    \hline
    Type \& Number & Recovered   & $\eta_\mathrm{1param}$ & $\eta_\mathrm{2param}$  \\
    of Injections & $\mathrm{FAR_{GW}}<864/\mathrm{d}$ &  $\left[>5\sigma\right]$ & $\left[>5\sigma\right]$  \\
    \hline
    \hline
    Dim1-KAM-LVD & 46.5\% & \cellcolor{magenta!30}\textbf{37.2\%} & \cellcolor{yellow!70}\textbf{44.2\%}   \\
    = 86 & = 40/86 & = 32/86 & = 38/86  \\
    \hline
    Dim3-KAM-LVD & 83.3\% & \cellcolor{magenta!30}\textbf{75.8\%} & \cellcolor{yellow!70}\textbf{81.5\%}   \\
     = 1386 & = 1154/1386 & = 1051/1386 & = 1130/1386  \\
     \hline
     Sch1-KAM-LVD & 39.1\% & \cellcolor{magenta!30}\textbf{30.4\%} & \cellcolor{yellow!70}\textbf{34.8\%}   \\
     = 23& = 9/23 & = 7/23 & = 8/23  \\
     \hline
     Sch2-KAM-LVD & 99.3\% & \cellcolor{magenta!30}\textbf{99.0\%} & \cellcolor{yellow!70}\textbf{99.2\%}   \\
     = 2329& = 2312/2329 & = 2305/2329 & = 2310/2329  \\
     \hline
     Sch3-KAM-LVD & 99.8\% & \cellcolor{magenta!30}\textbf{99.6\%} & \cellcolor{yellow!70}\textbf{99.7\%}   \\
     = 2398& = 2393/2398 & = 2388/2398 & = 2391/2398  \\
    \hline
    \end{tabular}
    }
    \end{table*}

\section{Single Detector Super-K}
\label{appC}

\begin{figure}[!ht]
    \centering
      \includegraphics[width=.7\linewidth]{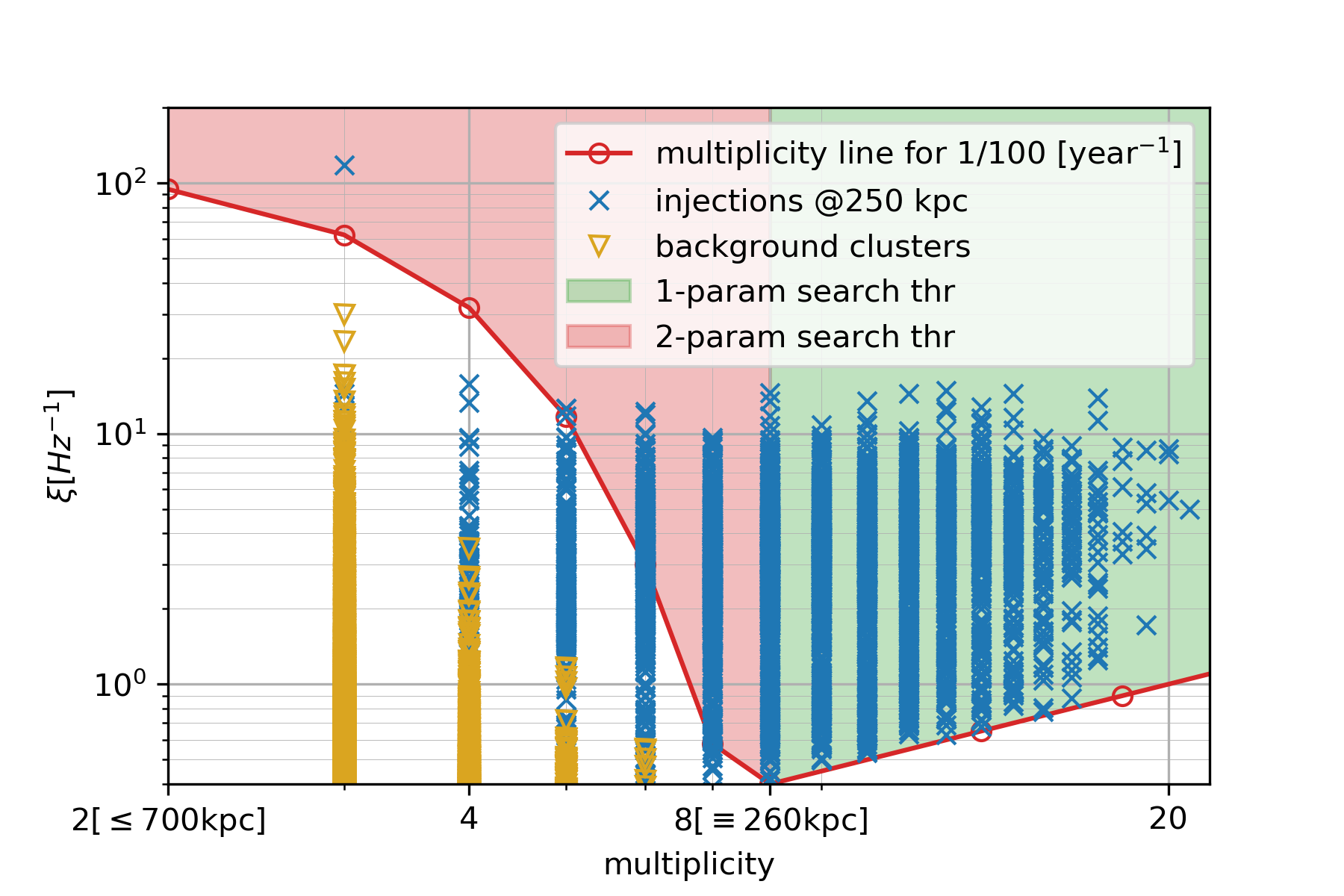}
      \caption{The $\xi$-multiplicity map for Super-K (as a leading example) with the simulated background and injections with SN1987A emission model at 250 kpc.}
      \label{fig:superk_250pc}
    \end{figure}

For the Super-K case, we do an analysis similar to that in Sec.~\ref{sec:single_det}. In this case, the Super-K detector is sensitive beyond the Small Magellanic Cloud. In order to show the improvement of our method we focus on a distance of 250 kpc. We provide the $\xi$ vs multiplicity plot in Fig. \ref{fig:superk_250pc} where the gain region is highlighted.

The efficiency comparison between the 1-parameter and the 2-parameter method for this distance is given in Tab.~\ref{tab:suk_single}. There are not many interesting objects for our target at 250-kpc distance; nevertheless, this exercise is done as a proof of concept of our method for the Super-K detector. Our method in fact could play a role for a future detector like Hyper-Kamiokande \cite{hyperK} to detect CCSNe in Andromeda galaxy and beyond.

    \begin{table}[!ht]
    \caption{Efficiency ($\eta$) comparison between 1-parameter and 2-parameter method of single detector Super-K with $D=250$ kpc for $\mathrm{FAR_{\nu}}\leq1/100\,\mathrm{[year^{-1}]}$}
    \label{tab:suk_single}
    \centering
    {\renewcommand{\arraystretch}{1.2}
    \begin{tabular}{|c  | c | c | c |} 
    \hline
    $D$  & Noise   & $\eta_\mathrm{1param}$ & $\eta_\mathrm{2param}$  \\
    
    [kpc]&  $\left[<1/100\,\mathrm{yr}\right]$ &  $\left[<1/100\,\mathrm{yr}\right]$ & $\left[<1/100\,\mathrm{yr}\right]$  \\
     \hline
    \hline
    \cellcolor{green!30}250  & 0/49200 & \cellcolor{magenta!30}2575/3645=\textbf{70.6\%}  & \cellcolor{yellow!70}3117/3645=\textbf{85.5\%}   \\
    \hline
    \end{tabular}
    }
    \end{table}

\bibliography{references}

\begin{thebibliography}{54}
\providecommand{\natexlab}[1]{#1}
\providecommand{\url}[1]{\texttt{#1}}
\expandafter\ifx\csname urlstyle\endcsname\relax
  \providecommand{\doi}[1]{doi: #1}\else
  \providecommand{\doi}{doi: \begingroup \urlstyle{rm}\Url}\fi

\bibitem[Pagliaroli et~al.(2009{\natexlab{a}})Pagliaroli, Vissani, Coccia, and
  Fulgione]{pagliaroli_PRL}
G.~Pagliaroli, F.~Vissani, E.~Coccia, and W.~Fulgione.
\newblock Neutrinos from supernovae as a trigger for gravitational wave search.
\newblock \emph{Phys. Rev. Lett.}, 103:\penalty0 031102, Jul
  2009{\natexlab{a}}.
\newblock \doi{10.1103/PhysRevLett.103.031102}.
\newblock URL \url{https://link.aps.org/doi/10.1103/PhysRevLett.103.031102}.

\bibitem[Leonor et~al.(2010)Leonor, Cadonati, Coccia,
  D{\textquotesingle}Antonio, Credico, Fafone, Frey, Fulgione, Katsavounidis,
  Ott, Pagliaroli, Scholberg, Thrane, and Vissani]{leonor}
I~Leonor, L~Cadonati, E~Coccia, S~D{\textquotesingle}Antonio, A~Di Credico,
  V~Fafone, R~Frey, W~Fulgione, E~Katsavounidis, C~D Ott, G~Pagliaroli,
  K~Scholberg, E~Thrane, and F~Vissani.
\newblock Searching for prompt signatures of nearby core-collapse supernovae by
  a joint analysis of neutrino and gravitational wave data.
\newblock \emph{Classical and Quantum Gravity}, 27\penalty0 (8):\penalty0
  084019, apr 2010.
\newblock \doi{10.1088/0264-9381/27/8/084019}.
\newblock URL \url{https://doi.org/10.1088%2F0264-9381%2F27%2F8%2F084019}.

\bibitem[Hirata et~al.(1987)]{hirata1987}
K.~Hirata et~al.
\newblock Observation of a neutrino burst from the supernova sn1987a.
\newblock \emph{Phys. Rev. Lett.}, 58:\penalty0 1490--1493, Apr 1987.
\newblock \doi{10.1103/PhysRevLett.58.1490}.
\newblock URL \url{https://link.aps.org/doi/10.1103/PhysRevLett.58.1490}.

\bibitem[Bionta et~al.(1987)]{Bionta1987}
R.~M. Bionta et~al.
\newblock Observation of a neutrino burst in coincidence with supernova 1987a
  in the large magellanic cloud.
\newblock \emph{Phys. Rev. Lett.}, 58:\penalty0 1494, 1987.
\newblock \doi{10.1103/PhysRevLett.58.1494}.

\bibitem[Alexeyev et~al.(1988)]{alexeyev}
E.N. Alexeyev et~al.
\newblock Detection of the neutrino signal from sn 1987a in the lmc using the
  inr baksan underground scintillation telescope.
\newblock \emph{Physics Letters B}, 205\penalty0 (2):\penalty0 209 -- 214,
  1988.
\newblock ISSN 0370-2693.
\newblock \doi{https://doi.org/10.1016/0370-2693(88)91651-6}.
\newblock URL
  \url{http://www.sciencedirect.com/science/article/pii/0370269388916516}.

\bibitem[Fukuda et~al.(2003)]{superK}
Y.~Fukuda et~al.
\newblock The super-kamiokande detector.
\newblock \emph{Nucl. Instrum. Meth.}, A501:\penalty0 418--462, 2003.
\newblock \doi{10.1016/S0168-9002(03)00425-X}.

\bibitem[Aglietta et~al.(1992)]{lvd_det}
M.~Aglietta et~al.
\newblock The most powerful scintillator supernovae detector: Lvd.
\newblock \emph{Il Nuovo Cimento A Series 11}, 105\penalty0 (12):\penalty0
  1793--1804, 1992.
\newblock \doi{10.1007/BF02740929}.
\newblock cited By 103.

\bibitem[Suekane et~al.(2004)]{kamland}
F.~Suekane et~al.
\newblock An overview of the kamland 1-kiloton liquid scintillator.
\newblock pages 279--290, 2004.
\newblock \href{https://arxiv.org/abs/physics/0404071}{arXiv:physics/0404071}.

\bibitem[Aartsen et~al.(2017)]{icecube}
M.G. Aartsen et~al.
\newblock The icecube neutrino observatory: instrumentation and online systems.
\newblock \emph{Journal of Instrumentation}, 12\penalty0 (03):\penalty0
  P03012--P03012, mar 2017.
\newblock \doi{10.1088/1748-0221/12/03/p03012}.
\newblock URL \url{https://doi.org/10.1088%2F1748-0221%2F12%2F03%2Fp03012}.

\bibitem[Antonioli et~al.(2004)]{Antonioli2004}
Pietro Antonioli et~al.
\newblock Snews: the supernova early warning system.
\newblock \emph{New Journal of Physics}, 6:\penalty0 114--114, sep 2004.
\newblock \doi{10.1088/1367-2630/6/1/114}.
\newblock URL \url{https://doi.org/10.1088%2F1367-2630%2F6%2F1%2F114}.

\bibitem[Kharusi et~al.(2021)]{Al_Kharusi_2021}
S~Al Kharusi et~al.
\newblock {SNEWS} 2.0: a next-generation supernova early warning system for
  multi-messenger astronomy.
\newblock \emph{New Journal of Physics}, 23\penalty0 (3):\penalty0 031201, mar
  2021.
\newblock \doi{10.1088/1367-2630/abde33}.
\newblock URL \url{https://doi.org/10.1088/1367-2630/abde33}.

\bibitem[Abbott et~al.(2017)]{Abbott2017}
B.~P. Abbott et~al.
\newblock Multi-messenger observations of a binary neutron star merger*.
\newblock \emph{The Astrophysical Journal}, 848\penalty0 (2):\penalty0 L12, oct
  2017.
\newblock \doi{10.3847/2041-8213/aa91c9}.
\newblock URL \url{https://doi.org/10.3847%2F2041-8213%2Faa91c9}.

\bibitem[Aasi et~al.(2015)]{ligo}
J.~Aasi et~al.
\newblock {Advanced LIGO}.
\newblock \emph{Class. Quant. Grav.}, 32:\penalty0 074001, 2015.
\newblock \doi{10.1088/0264-9381/32/7/074001}.

\bibitem[Acernese et~al.(2015)]{virgo}
F.~Acernese et~al.
\newblock {Advanced Virgo: a second-generation interferometric gravitational
  wave detector}.
\newblock \emph{Class. Quant. Grav.}, 32\penalty0 (2):\penalty0 024001, 2015.
\newblock \doi{10.1088/0264-9381/32/2/024001}.

\bibitem[Akutsu et~al.(2020)]{kagra}
T.~Akutsu et~al.
\newblock Overview of kagra: Detector design and construction history, 2020.
\newblock \href{https://arxiv.org/abs/2005.05574}{arXiv:2005.05574}.

\bibitem[Ott(2009)]{Ott_2009}
Christian~D Ott.
\newblock The gravitational-wave signature of core-collapse supernovae.
\newblock \emph{Classical and Quantum Gravity}, 26\penalty0 (6):\penalty0
  063001, feb 2009.
\newblock \doi{10.1088/0264-9381/26/6/063001}.
\newblock URL \url{https://doi.org/10.1088%2F0264-9381%2F26%2F6%2F063001}.

\bibitem[Abdikamalov et~al.(2020)Abdikamalov, Pagliaroli, and
  Radice]{Abdikamalov:2020jzn}
Ernazar Abdikamalov, Giulia Pagliaroli, and David Radice.
\newblock Gravitational waves from core-collapse supernovae, 2020.
\newblock \href{https://arxiv.org/abs/2010.04356}{arXiv:2010.04356}.

\bibitem[Powell and M\"uller(2020)]{powell2020}
Jade Powell and Bernhard M\"uller.
\newblock {Three-dimensional core-collapse supernova simulations of massive and
  rotating progenitors}.
\newblock \emph{Mon. Not. Roy. Astron. Soc.}, 494\penalty0 (4):\penalty0
  4665--4675, 2020.
\newblock \doi{10.1093/mnras/staa1048}.

\bibitem[Szczepanczyk et~al.(2021)Szczepanczyk, Antelis, Benjamin, Cavaglia,
  Gondek-Rosinska, Hansen, Klimenko, Morales, Moreno, Mukherjee, Nurbek,
  Powell, Singh, Sitmukhambetov, Szewczyk, Westhouse, Valdez, Vedovato, Zheng,
  and Zanolin]{Szczepanczyk}
Marek Szczepanczyk, Javier Antelis, Michael Benjamin, Marco Cavaglia, Dorota
  Gondek-Rosinska, Travis Hansen, Sergey Klimenko, Manuel Morales, Claudia
  Moreno, Soma Mukherjee, Gaukhar Nurbek, Jade Powell, Neha Singh, Satzhan
  Sitmukhambetov, Pawel Szewczyk, Jonathan Westhouse, Oscar Valdez, Gabriele
  Vedovato, Yanyan Zheng, and Michele Zanolin.
\newblock Detecting and reconstructing gravitational waves from the next
  galactic core-collapse supernova in the advanced detector era, 2021.
\newblock \href{https://arxiv.org/abs/2104.06462}{arXiv:2104.06462}.

\bibitem[Halim(2020)]{halimtesi}
O.~Halim.
\newblock \emph{Searching for Core-Collapse Supernovae in the Multimessenger
  Era: Low Energy Neutrinos and Gravitational Waves}.
\newblock PhD thesis, Gran Sasso Science Institute (GSSI), 2020.

\bibitem[Halim et~al.(2020)]{halim2019}
Odysse Halim et~al.
\newblock Expanding core-collapse supernova search horizon of neutrino
  detectors.
\newblock \emph{Journal of Physics: Conference Series}, 1468:\penalty0 012154,
  feb 2020.
\newblock \doi{10.1088/1742-6596/1468/1/012154}.
\newblock URL \url{https://doi.org/10.1088%2F1742-6596%2F1468%2F1%2F012154}.

\bibitem[Aso et~al.(2008)]{Aso_2008}
Yoichi Aso et~al.
\newblock Search method for coincident events from {LIGO} and {IceCube}
  detectors.
\newblock \emph{Classical and Quantum Gravity}, 25\penalty0 (11):\penalty0
  114039, may 2008.
\newblock \doi{10.1088/0264-9381/25/11/114039}.
\newblock URL \url{https://doi.org/10.1088%2F0264-9381%2F25%2F11%2F114039}.

\bibitem[Baret et~al.(2012)]{multimess}
Bruny Baret et~al.
\newblock Multimessenger science reach and analysis method for common sources
  of gravitational waves and high-energy neutrinos.
\newblock \emph{Phys. Rev. D}, 85:\penalty0 103004, 2012.
\newblock \doi{10.1103/PhysRevD.85.103004}.

\bibitem[Adri{\'{a}}n-Mart{\'{\i}}nez et~al.(2013)]{Adrian2013}
S~Adri{\'{a}}n-Mart{\'{\i}}nez et~al.
\newblock A first search for coincident gravitational waves and high energy
  neutrinos using ligo, virgo and antares data from 2007.
\newblock \emph{Journal of Cosmology and Astroparticle Physics}, 2013\penalty0
  (06):\penalty0 008--008, jun 2013.
\newblock \doi{10.1088/1475-7516/2013/06/008}.
\newblock URL \url{https://doi.org/10.1088%2F1475-7516%2F2013%2F06%2F008}.

\bibitem[Palma(2014)]{dipalma2014}
Irene~Di Palma.
\newblock Multimessenger astrophysics: When gravitational waves meet high
  energy neutrinos.
\newblock \emph{Nuclear Instruments and Methods in Physics Research Section A:
  Accelerators, Spectrometers, Detectors and Associated Equipment},
  742:\penalty0 124 -- 129, 2014.
\newblock ISSN 0168-9002.
\newblock \doi{https://doi.org/10.1016/j.nima.2013.10.076}.
\newblock URL
  \url{http://www.sciencedirect.com/science/article/pii/S0168900213014836}.
\newblock 4th Roma International Conference on Astroparticle Physics.

\bibitem[Abbott et~al.(2008)]{gwgrbabbott}
B.~Abbott et~al.
\newblock Search for gravitational waves associated with 39 gamma-ray bursts
  using data from the second, third, and fourth ligo runs.
\newblock \emph{Phys. Rev. D}, 77:\penalty0 062004, Mar 2008.
\newblock \doi{10.1103/PhysRevD.77.062004}.
\newblock URL \url{https://link.aps.org/doi/10.1103/PhysRevD.77.062004}.

\bibitem[Ashton et~al.(2018)]{Ashton_2018}
G.~Ashton et~al.
\newblock Coincident detection significance in multimessenger astronomy.
\newblock \emph{The Astrophysical Journal}, 860\penalty0 (1):\penalty0 6, jun
  2018.
\newblock \doi{10.3847/1538-4357/aabfd2}.
\newblock URL \url{https://doi.org/10.3847%2F1538-4357%2Faabfd2}.

\bibitem[Klimenko and Mitselmakher(2004)]{Klimenko_2004}
S~Klimenko and G~Mitselmakher.
\newblock A wavelet method for detection of gravitational wave bursts.
\newblock \emph{Classical and Quantum Gravity}, 21\penalty0 (20):\penalty0
  S1819--S1830, sep 2004.
\newblock \doi{10.1088/0264-9381/21/20/025}.
\newblock URL \url{https://doi.org/10.1088%2F0264-9381%2F21%2F20%2F025}.

\bibitem[Klimenko et~al.(2008)]{Klimenko2008}
S~Klimenko et~al.
\newblock A coherent method for detection of gravitational wave bursts.
\newblock \emph{Classical and Quantum Gravity}, 25\penalty0 (11):\penalty0
  114029, may 2008.
\newblock \doi{10.1088/0264-9381/25/11/114029}.
\newblock URL \url{https://doi.org/10.1088%2F0264-9381%2F25%2F11%2F114029}.

\bibitem[Drago(2010)]{dragotesi}
M.~Drago.
\newblock \emph{Search for Transient Gravitational Wave Signals with unknown
  Waveform in the LIGO Virgo Network of Interferometric Detectors using a fully
  Coherent Algorithm}.
\newblock PhD thesis, Universit{\`a} degli Studi di Padova, 2010.

\bibitem[Necula et~al.(2012)Necula, Klimenko, and Mitselmakher]{Necula_2012}
V~Necula, S~Klimenko, and G~Mitselmakher.
\newblock Transient analysis with fast wilson-daubechies time-frequency
  transform, jun 2012.
\newblock URL \url{https://doi.org/10.1088/1742-6596/363/1/012032}.

\bibitem[Abbott et~al.(2016{\natexlab{a}})]{gw150914}
B.~P. Abbott et~al.
\newblock Observation of gravitational waves from a binary black hole merger.
\newblock \emph{Phys. Rev. Lett.}, 116:\penalty0 061102, Feb
  2016{\natexlab{a}}.
\newblock \doi{10.1103/PhysRevLett.116.061102}.
\newblock URL \url{https://link.aps.org/doi/10.1103/PhysRevLett.116.061102}.

\bibitem[Abbott et~al.(2020{\natexlab{a}})]{em_gw_ccsne}
B.~P. Abbott et~al.
\newblock Optically targeted search for gravitational waves emitted by
  core-collapse supernovae during the first and second observing runs of
  advanced ligo and advanced virgo.
\newblock \emph{Phys. Rev. D}, 101:\penalty0 084002, Apr 2020{\natexlab{a}}.
\newblock \doi{10.1103/PhysRevD.101.084002}.
\newblock URL \url{https://link.aps.org/doi/10.1103/PhysRevD.101.084002}.

\bibitem[Janka(2012)]{jankarev}
Hans-Thomas Janka.
\newblock Explosion mechanisms of core-collapse supernovae.
\newblock \emph{Annual Review of Nuclear and Particle Science}, 62\penalty0
  (1):\penalty0 407--451, 2012.
\newblock \doi{10.1146/annurev-nucl-102711-094901}.

\bibitem[O'Connor and Ott(2011)]{OConnor:2010moj}
Evan O'Connor and Christian~D. Ott.
\newblock {Black Hole Formation in Failing Core-Collapse Supernovae}.
\newblock \emph{Astrophys. J.}, 730:\penalty0 70, 2011.
\newblock \doi{10.1088/0004-637X/730/2/70}.

\bibitem[Radice et~al.(2019)Radice, Morozova, Burrows, Vartanyan, and
  Nagakura]{radice}
David Radice, Viktoriya Morozova, Adam Burrows, David Vartanyan, and Hiroki
  Nagakura.
\newblock {Characterizing the Gravitational Wave Signal from Core-Collapse
  Supernovae}.
\newblock \emph{Astrophys. J. Lett.}, 876\penalty0 (1):\penalty0 L9, 2019.
\newblock \doi{10.3847/2041-8213/ab191a}.

\bibitem[Dimmelmeier et~al.(2008)]{dimmelmeier2008}
Harald Dimmelmeier et~al.
\newblock Gravitational wave burst signal from core collapse of rotating stars.
\newblock \emph{Phys. Rev. D}, 78:\penalty0 064056, Sep 2008.
\newblock \doi{10.1103/PhysRevD.78.064056}.
\newblock URL \url{https://link.aps.org/doi/10.1103/PhysRevD.78.064056}.

\bibitem[Scheidegger et~al.(2010)Scheidegger, Kaeppeli, Whitehouse, Fischer,
  and Liebendoerfer]{Scheidegger:2010en}
S.~Scheidegger, R.~Kaeppeli, S.~C. Whitehouse, T.~Fischer, and
  M.~Liebendoerfer.
\newblock {The Influence of Model Parameters on the Prediction of Gravitational
  wave Signals from Stellar Core Collapse}.
\newblock \emph{Astron. Astrophys.}, 514:\penalty0 A51, 2010.
\newblock \doi{10.1051/0004-6361/200913220}.

\bibitem[Woosley and Heger(2006)]{woosley}
Stan Woosley and Alexander Heger.
\newblock {The Progenitor stars of gamma-ray bursts}.
\newblock \emph{Astrophys. J.}, 637:\penalty0 914--921, 2006.
\newblock \doi{10.1086/498500}.

\bibitem[H{\"u}depohl(2014)]{hudepohl}
L.~H{\"u}depohl.
\newblock \emph{Neutrinos from the Formation, Cooling and Black Hole Collapse
  of Neutron Stars}.
\newblock PhD thesis, Technische Universit{\"a}t M{\"u}nchen, 2014.

\bibitem[Pagliaroli et~al.(2009{\natexlab{b}})]{pagliaroli2009}
G.~Pagliaroli et~al.
\newblock Improved analysis of sn1987a antineutrino events.
\newblock \emph{Astroparticle Physics}, 31\penalty0 (3):\penalty0 163 -- 176,
  2009{\natexlab{b}}.
\newblock ISSN 0927-6505.
\newblock \doi{https://doi.org/10.1016/j.astropartphys.2008.12.010}.
\newblock URL
  \url{http://www.sciencedirect.com/science/article/pii/S0927650508001965}.

\bibitem[Vissani et~al.(2011)Vissani, Pagliaroli, and
  Costantini]{pagliaroli_ccsn}
Francesco Vissani, Giulia Pagliaroli, and Maria~Laura Costantini.
\newblock {A parameterized model for supernova electron antineutrino emission
  and its applications}.
\newblock \emph{J. Phys. Conf. Ser.}, 309:\penalty0 012025, 2011.
\newblock \doi{10.1088/1742-6596/309/1/012025}.

\bibitem[Abbott et~al.(2016{\natexlab{b}})]{SNTargeted2016}
B.~P. Abbott et~al.
\newblock First targeted search for gravitational-wave bursts from
  core-collapse supernovae in data of first-generation laser interferometer
  detectors.
\newblock \emph{Phys. Rev. D}, 94:\penalty0 102001, Nov 2016{\natexlab{b}}.
\newblock \doi{10.1103/PhysRevD.94.102001}.
\newblock URL \url{https://link.aps.org/doi/10.1103/PhysRevD.94.102001}.

\bibitem[Klimenko et~al.(2016)Klimenko, Vedovato, Drago, Salemi, Tiwari, Prodi,
  Lazzaro, Ackley, Tiwari, Da~Silva, and Mitselmakher]{Klimenko:2015ypf}
S.~Klimenko, G.~Vedovato, M.~Drago, F.~Salemi, V.~Tiwari, G.~A. Prodi,
  C.~Lazzaro, K.~Ackley, S.~Tiwari, C.~F. Da~Silva, and G.~Mitselmakher.
\newblock {Method for detection and reconstruction of gravitational wave
  transients with networks of advanced detectors}.
\newblock \emph{Phys. Rev. D}, 93\penalty0 (4):\penalty0 042004, 2016.
\newblock \doi{10.1103/PhysRevD.93.042004}.

\bibitem[Drago et~al.(2021)Drago, Gayathri, Klimenko, Lazzaro, Milotti,
  Mitselmakher, Necula, O'Brian, Prodi, Salemi, Szczepanczyk, Tiwari, Tiwari,
  Vedovato, and Yakushin]{Drago:2020kic}
M.~Drago, V.~Gayathri, S.~Klimenko, C.~Lazzaro, E.~Milotti, G.~Mitselmakher,
  V.~Necula, B.~O'Brian, G.~A. Prodi, F.~Salemi, M.~Szczepanczyk, S.~Tiwari,
  V.~Tiwari, G.~Vedovato, and I.~Yakushin.
\newblock Coherent waveburst, a pipeline for unmodeled gravitational-wave data
  analysis, 2021.
\newblock \href{https://arxiv.org/abs/2006.12604}{arXiv:2006.12604}.

\bibitem[Abbott et~al.(2020{\natexlab{b}})]{Abbott:2020qfu}
B.~P. Abbott et~al.
\newblock {Prospects for observing and localizing gravitational-wave transients
  with Advanced LIGO, Advanced Virgo and KAGRA}.
\newblock \emph{Living Rev. Rel.}, 23\penalty0 (1):\penalty0 3,
  2020{\natexlab{b}}.
\newblock \doi{10.1007/s41114-020-00026-9}.

\bibitem[Szczepa\'nczyk(2018)]{marektesi}
Marek~J. Szczepa\'nczyk.
\newblock \emph{Multimessenger Astronomy with Gravitational Waves from
  Core-Collapse Supernovae}.
\newblock PhD thesis, Embry-Riddle Aeronautical University, 2018.

\bibitem[Agafonova et~al.(2015)]{Agafonova2014}
N.Y. Agafonova et~al.
\newblock Implication for the core-collapse supernova rate from 21 years of
  data of the large volume detector.
\newblock \emph{Astrophys.\ J.}, 802\penalty0 (1):\penalty0 47, 2015.
\newblock \doi{10.1088/0004-637X/802/1/47}.

\bibitem[Ikeda et~al.(2007)]{Ikeda2007}
M.~Ikeda et~al.
\newblock Search for supernova neutrino bursts at super-kamiokande.
\newblock \emph{Astrophys. J.}, 669:\penalty0 519--524, 2007.
\newblock \doi{10.1086/521547}.

\bibitem[Abe et~al.(2016)]{Abe2016}
K.~Abe et~al.
\newblock Real-time supernova neutrino burst monitor at super-kamiokande.
\newblock \emph{Astropart. Phys.}, 81:\penalty0 39--48, 2016.
\newblock \doi{10.1016/j.astropartphys.2016.04.003}.

\bibitem[Casentini et~al.(2018)]{Casentini2018}
C.~Casentini et~al.
\newblock Pinpointing astrophysical bursts of low-energy neutrinos embedded
  into the noise.
\newblock \emph{JCAP}, 1808\penalty0 (08):\penalty0 010, 2018.
\newblock \doi{10.1088/1475-7516/2018/08/010}.

\bibitem[Mattiazzi et~al.(2021)Mattiazzi, Lamoureux, and
  Collazuol]{Mattiazzi:2021zcb}
Marco Mattiazzi, Mathieu Lamoureux, and Gianmaria Collazuol.
\newblock On a new statistical technique for the real-time recognition of
  ultra-low multiplicity astrophysical neutrino burst, 2021.
\newblock \href{https://arxiv.org/abs/2106.12345}{arXiv:2106.12345}.

\bibitem[Eguchi et~al.(2003)]{kam_bg}
K.~Eguchi et~al.
\newblock First results from kamland: Evidence for reactor antineutrino
  disappearance.
\newblock \emph{Phys. Rev. Lett.}, 90:\penalty0 021802, Jan 2003.
\newblock \doi{10.1103/PhysRevLett.90.021802}.
\newblock URL \url{https://link.aps.org/doi/10.1103/PhysRevLett.90.021802}.

\bibitem[Abe et~al.(2018)]{hyperK}
K.~Abe et~al.
\newblock Hyper-kamiokande design report, 2018.
\newblock \href{https://arxiv.org/abs/1805.04163}{arXiv:1805.04163}.

\end{thebibliography}

\bibliographystyle{unsrtnat}

\end{document}